\documentclass[12pt,draftclsnofoot,onecolumn]{IEEEtran}

\usepackage{amsmath, mathrsfs,balance}
\usepackage{bm}

\usepackage{graphicx}
\usepackage{color}
\usepackage{tikz}
\usepackage{pgfplots}
\pgfplotsset{compat=newest}
\usetikzlibrary{patterns}
\usetikzlibrary{positioning}
\usetikzlibrary{datavisualization}
\usetikzlibrary{datavisualization.formats.functions}
\usetikzlibrary{backgrounds}
\usetikzlibrary{shapes,snakes}

\makeatletter
\def\maketag@@@#1{\hbox{\m@th\normalfont\normalsize#1}}
\makeatother

\usepackage[none]{hyphenat}

\newcommand{\subparagraph}{}
\usepackage[compact]{titlesec}
\titlespacing*{\section}{2pt}{1\baselineskip}{0.9\baselineskip}

\allowdisplaybreaks

\usepackage{amsfonts}
\usepackage{amssymb}
\usepackage{color}
\usepackage{multirow}
\usepackage{graphicx}

\usepackage{caption}
\usepackage{subcaption}
\usepackage[numbers,sort&compress]{natbib}
\usepackage{balance}

\usepackage{mathbbol}

\def\mindex#1{\index{#1}}

\def\sq{\hbox{\rlap{$\sqcap$}$\sqcup$}}
\def\qed{\ifmmode\sq\else{\unskip\nobreak\hfil
\penalty50\hskip1em\null\nobreak\hfil\sq
\parfillskip=0pt\finalhyphendemerits=0\endgraf}\fi\medskip}

\long\def\defbox#1{\framebox[.9\hsize][c]{\parbox{.85\hsize}{%
\parindent=0pt
\baselineskip=12pt plus .1pt      
\parskip=6pt plus 1.5pt minus 1pt 
 #1}}}

\long\def\beginbox#1\endbox{\subsection*{}%
\hbox{\hspace{.05\hsize}\defbox{\medskip#1\bigskip}}%
\subsection*{}}

\def\endbox{}

\newsavebox{\junk}
\savebox{\junk}[1.6mm]{\hbox{$|\!|\!|$}}

\def\argmin{\mathop{\rm arg\, min}}

\newcommand{\field}[1]{\mathbb{#1}}

\def\Re{\field{R}}

\def\intgr{\field{Z}}

\def\bC{{\mathbb C}}

\def\bE{{\mathbb E}}

\def\bR{{\mathbb R}}

\def\bV{{\mathbb V}}

\def\bfA{{\bf A}}
\def\bfB{{\bf B}}
\def\bfC{{\bf C}}

\def\bfF{{\bf F}}

\def\bfH{{\bf H}}
\def\bfI{{\bf I}}

\def\bfa{{\bf a}}
\def\bfb{{\bf b}}
\def\bfc{{\bf c}}
\def\bfd{{\bf d}}

\def\bfg{{\bf g}}
\def\bfh{{\bf h}}

\def\bfq{{\bf q}}
\def\bfr{{\bf r}}

\def\bfu{{\bf u}}
\def\bfv{{\bf v}}
\def\bfw{{\bf w}}
\def\bfx{{\bf x}}

\def\ttd{{\mathtt d}}

\def\tti{{\mathtt i}}

\def\ttm{{\mathtt m}}

\def\tto{{\mathtt o}}

\def\ttt{{\mathtt t}}

\def\sfH{{\sf H}}

\def\sfa{{\sf a}}

\def\sfr{{\sf r}}

\def\bfmath#1{{\mathchoice{\mbox{\boldmath$#1$}}%
{\mbox{\boldmath$#1$}}%
{\mbox{\boldmath$\scriptstyle#1$}}%
{\mbox{\boldmath$\scriptscriptstyle#1$}}}}

\def\bfmY{\bfmath{Y}}

\def\bfmhhaY{\bfmath{\hhaY}} 
\def\bfmhhaY{\hbox to 0pt{$\widehat{\bfmY}$\hss}\widehat{\phantom{\raise 1.25pt\hbox{$\bfmY$}}}}

\def\til={{\widetilde =}}

\def\clA{{\cal A}}

\def\clC{{\cal C}}

\def\clN{{\cal N}}

\def\clU{{\cal U}}
\def\clV{{\cal V}}

 \def\FRAC#1#2#3{\genfrac{}{}{}{#1}{#2}{#3}}

\def\ddtp{{\mathchoice{\FRAC{1}{d^{\hbox to 2pt{\rm\tiny +\hss}}}{dt}}%
{\FRAC{1}{d^{\hbox to 2pt{\rm\tiny +\hss}}}{dt}}%
{\FRAC{3}{d^{\hbox to 2pt{\rm\tiny +\hss}}}{dt}}%
{\FRAC{3}{d^{\hbox to 2pt{\rm\tiny +\hss}}}{dt}}}}

\def\average#1,#2,{{1\over #2} \sum_{#1}^{#2}}

\def\eye(#1){{\bf(#1)}\quad}

\def\var{{\bV\sfa\sfr}}

\def\eq#1/{(\ref{e:#1})}

\newcommand{\beqn}[1]{\notes{#1}%
\begin{eqnarray} \elabel{#1}}

\newcommand{\eeqn}{\end{eqnarray} }

\newcommand{\beq}[1]{\notes{#1}%
\begin{equation}\elabel{#1}}

\newcommand{\eeq}{\end{equation}}

\def\bdes{\begin{description}}
\def\edes{\end{description}}

\newcounter{rmnum}

\newcounter{anum}

{\end{list}}

\def\ass(#1:#2){(#1\ref{#1:#2})}

\def\ritem#1{
\item[{\sf \ass(\current_model:#1)}]
}

\newenvironment{recall-ass}[1]{%
\begin{description}
\def\current_model{#1}}{
\end{description}
}

\long\def\comment#1{}

\newcommand{\Fm}{{\bf F}}

\newcommand{\Uc}{{\cal U}}

\newcommand{\Vc}{{\cal V}}

\newcommand{\Gammam}{\boldsymbol{\Gamma}}

\renewcommand{\Re}{{\rm Re}}

\newcommand{\transp}{{\sf T}}
\renewcommand{\vec}{{\rm vec}}

\def\snrbef{{\mathsf{SNR}_\text{BBF}}}

\usepackage{tikz}
\usepackage{pgfplots}
\pgfplotsset{compat=newest}
\usetikzlibrary{patterns}
\usetikzlibrary{positioning}
\usetikzlibrary{datavisualization}
\usetikzlibrary{datavisualization.formats.functions}
\usetikzlibrary{backgrounds}
\usetikzlibrary{shapes,snakes}
\usepackage{textcomp}
\usetikzlibrary{automata}

\allowdisplaybreaks

\usepackage{algorithm}
\usepackage{algpseudocode}
\usepackage{pifont}
\usepackage{varwidth}

\def\one{{\bf 1}}

\usepackage{multicol,lipsum}

\def\herm{{\sfH}}

\def\snr{{\mathsf{SNR}}}

\def\ptot{{P_{\ttt \tto \ttt}}}
\def\pdim{{P_{\ttd \tti \ttm}}}

\def\cg{{\clC\clN}}
\def\matlab{{MATLAB\textcopyright\,}}

\usepackage{color}
\newcommand{\figref}[1]{Fig.~\ref{#1}}
\usetikzlibrary{decorations.markings}

\begin{document}

\title{Efficient Beam Alignment for mmWave Single-Carrier Systems with Hybrid MIMO Transceivers}
\author{Xiaoshen Song, \IEEEmembership{Student Member, IEEE,} Saeid Haghighatshoar,  \IEEEmembership{Member, IEEE,} Giuseppe Caire, \IEEEmembership{Fellow, IEEE}}

\maketitle

\begin{abstract}
Communication at {\em millimeter wave} (mmWave) bands  is expected to become a key ingredient of next generation (5G) wireless networks.
Effective mmWave communications require fast and reliable methods for beamforming at both the {\em User Equipment} (UE) and the {\em Base Station} (BS) sides, in order to
achieve a sufficiently large {\em Signal-to-Noise Ratio} (SNR) after beamforming.
We refer to the problem of finding a pair of strongly coupled narrow beams at the transmitter and receiver as the {\em Beam Alignment} (BA) problem.
In this paper, we propose an efficient BA scheme for single-carrier mmWave communications. In the proposed scheme, the
BS periodically probes the channel in the downlink via a pre-specified pseudo-random beamforming codebook
and pseudo-random spreading codes, letting each UE estimate the
{\em Angle-of-Arrival / Angle-of-Departure} (AoA-AoD)
pair of the multipath channel for which the energy transfer is maximum.
We leverage the sparse nature of mmWave channels in the AoA-AoD domain to formulate the BA problem as the 
estimation of a sparse non-negative vector. 
Based on the recently developed {\em Non-Negative Least Squares} (NNLS) technique, we efficiently find the strongest AoA-AoD pair connecting each UE to the BS. We evaluate the  performance of the proposed scheme under a realistic channel model, where the propagation channel consists of a few 
multipath scattering components each having different delays, AoAs-AoDs, and Doppler shifts.
The channel model parameters are consistent with experimental channel measurements.  
Simulation results indicate that the proposed method is highly robust to fast channel variations
caused by the large Doppler spread between the multipath components. Furthermore, we also show that after achieving BA the beamformed 
channel is essentially frequency-flat, such that single-carrier communication needs no equalization in the time domain. 
\end{abstract}	


\section{Introduction}\label{introduction}

The majority of existing wireless communication systems operate in the sub-$6$ GHz microwave spectrum, which has now become very crowded.
As a result, {\em millimeter wave} (mmWave) spectrum  ranging from $30$ to $300$ GHz has been considered as an alternative to achieve very high data rates in
the next generation wireless systems.  At these frequencies, a signal bandwidth of 1GHz with {\em Signal-to-Noise Ratio} (SNR)  between 0dB and 3dB yields data rates $\sim 1$Gb/s per data stream.
A mmWave {\em Base Station} (BS) supporting multiple data streams  through the use of 
multiuser {\em Multiple-Input Multiple-Output} (MIMO)  can achieve tens of Gb/s of aggregate rate, 
thus fulfilling the requirements of enhanced Mobile Broad Band (eMBB) in 5G \cite{boccardi2014five,andrews2014will}.

A main challenge of communication at mmWaves is the very short range of isotropic propagation.
According to Frii's Law \cite{Zhaojie2016BA}, the effective area of an isotropic antenna decreases polynomially with frequency, therefore, the isotropic pathloss
at mmWaves is considerably larger compared with sub-$6$ GHz counterpart.
Moreover, signal propagation through scattering elements also suffers from a large attenuation at high frequencies.
Fortunately, the small wavelength of mmWave signals enables to pack a large number of antenna elements
in a small form factor, such that one can cope with the severe isotropic pathloss by using large antenna arrays both at the BS side and the {\em User Equipment} (UE) side, providing an overall large beamforming gain.   An essential component to obtain such large antenna gains consists of identifying  
suitable narrow beam combinations, i.e., a pair of {\em Angle of Departure} (AoD) at the BS and {\em Angle of Arrival} (AoA) at the UE, 
yielding a sufficiently large beamforming gain through the scatterers in the channel.
\footnote{We refer to AoD for the BS and AoA for the UE since the proposed scheme consists of downlink probing from the BS to the UEs.	Of course, due to the propagation angle reciprocity, the role of AoA and AoD is referred in the uplink.}
The problem of finding an AoA-AoD pair with a large channel gain is referred to as
{\em Initial Beam Training, Acquisition, or Alignment} in the literature (see references in Section \ref{related-work}).
Consistently with our previous work \cite{sxsBA2017}, we shall refer to it simply as {\em Beam Alignment} (BA).

It is important to note under which conditions the BA operation must be performed. In this work, we
focus on MIMO devices with a {\em Hybrid Digital Analog} (HDA)  structure.
HDA MIMO is widely proposed especially for mmWave systems, since the size and power consumption of all-digital architectures prevent the integration of many antenna elements on a small space. In a HDA implementation, the signal processing is done via the concatenation of an analog part implementing the beamforming functions, and a digital part implementing the baseband processing \cite{molisch2017hybrid,OverviewHeath2016}.
This poses some specific challenges: {i)  The signal received at the antennas passes through an analog beamforming network with only a limited number
of {\em Radio Frequency} (RF) chains, much smaller than the number of antennas.
Hence, the baseband signal processing has access to only a low-dimension projection of the whole received signal;
{ii) Due to the large isotropic pathloss, the received signal power is very low before beamforming, i.e., at every antenna port. Therefore, 
the implementation of BA is confronted with a very low SNR;
{iii) Because of the large number of antennas at both sides, the size of the channel matrix between each UE and the BS is very large. However,
extensive channel measurements have shown that mmWave channels typically exhibit an average of up to $3$ multipath components, each corresponding to
a scattering cluster with small delay/angle spreading \cite{Rappaport2014Capacity,Sayeed2014Sparse}. 
As a result, a suitable BA  scheme requires identifying a very sparse
set of AoA-AoDs in a very low-dimension  channel matrix \cite{RobertSOMP2017, Rappaport2017lowrank}.

The other fundamental aspect to the BA problem is that this is the first operation that a UE must accomplish in order to communicate with the BS. Hence,
while coarse frame and carrier frequency synchronization may be assumed (especially for the non-stand alone system, assisted by some other existing
cell operating at lower frequencies), the fine timing and Doppler shift compensation cannot be assumed. It follows that the BA operation must cope with significant timing offsets and
Doppler shifts. In addition, in a multpath propagation environment with paths coming from different directions, each path may be affected by a different Doppler shift.
In multicarrier (OFDM-based) systems, this may lead to significant inter-carrier interference, which has been typically ignored in 
most of the current literature.

\subsection{Related Work}\label{related-work}

The most straightforward BA method is an exhaustive search, where the BS and the UE scan all the AoA-AoD beam pairs until they find a strong one \cite{Rappaport2014Capacity}. This  is, however, prohibitively time-consuming, especially considering the very large dimension of the channel matrix due to very large number of antennas.
Several BA algorithms have been recently proposed in the literature. All these algorithms, in some way,
aim at achieving reliable BA while using less overhead than the exhaustive scheme.

In \cite{Palacios2017PseudoExhaus}, a two-stage pseudo-exhaustive BA scheme was proposed, where in the first stage,
the BS isotropically probes the channel, while the UE performs beam sweeping to find the best AoA, and in the second stage, the UE probes the channel along
the estimated AoA from the first stage, while the BS performs beam sweeping to find the best AoD.
A main limitation of \cite{Palacios2017PseudoExhaus} is that, due to the isotropic BS beamforming at the first stage, the scheme suffers from a very low pre-beamforming SNR \cite{Ghosh2013Backhaul, SaeidBA2016,RobertSOMP2017}, which may impairs the whole BA performance. 

Some mmWave standards such as IEEE 802.11ad \cite{80211ad} proposed to use multi-level hierarchical BA schemes (e.g., see also
 \cite{alkhateeb2014channel,Branka2016Overlap,Noh2017bisection,Hussain2017bisection}).
 The underlying idea is to start with sectors of wide beams to do a coarse BA and then shrink the beamwidth adaptively and successively to obtain a more refined BA.
The drawback of such schemes, however, is that each UE has its own specific AoA as seen from the BS side, thus, the BS needs to interact with each  UE individually. As a result, all these
hierarchical schemes require a coordination among the UEs and the BS, which is difficult to have at the initial channel acquisition stage.
Moreover, since hierarchical schemes requires interactive uplink-downlink communication 
between the BS and each individual UE, it is not clear how  the overhead of such schemes  scales in small cell scenarios 
with significant mobility of users across cells, where the BA procedure should be repeated at each handover.

The sparse nature of mmWave channels, i.e., large-dimension channel matrices along with very sparse scatterers in the AoA-AoD domain \cite{Rappaport2014Capacity,Sayeed2014Sparse}, motivates the application of  {\em Compressed Sensing} (CS) methods 
to speed up the BA. There are two groups of CS-based methods in the literature.
The first group (e.g., see \cite{AhmedFreqOMP2015, RobertSOMP2017,Time2017,AlkhateebTimeDomain2017}) applies 
CS to estimate the complex baseband channel coefficients. These algorithms are efficient and particularly attractive for multiuser scenarios, 
but they are based on the assumption that the instantaneous channel remains invariant during the whole
probing/measuring stage. As anticipated before, this assumption is difficult to meet at mmWaves because of the large Doppler spread between the 
multipath components coming from different angles, implying significant time-variations of the channel coefficients even for UEs with small mobility \cite{WeilerMeasure2014,HeathVariation2017,Rappaport2017lowrank}.\footnote{Notice that the channel delay spread and time-variation are greatly reduced {\em after BA is achieved}, since once the beams are aligned, the communication occurs only through a single multipath component
	with small effective angular spread, whose delay and Doppler shift can be well compensated  \cite{HeathVariation2017}.
	However, {\em before BA is achieved} the channel delay spread and time-variation can be large due to the presence of
	several mulipath components, each with its own delay and Doppler shift.
	In this case, even a small motion of a few centimeters traverses several wavelengths, potentially producing multiple
	deep fades \cite{WeilerMeasure2014}.}
The second group of CS-based schemes focuses on estimating the second-order statistics of the channel, i.e., the covariance of the channel matrix,
which is very robust to channel variations. In \cite{Rappaport2017lowrank} for example, a {\em Maximum Likelihood} (ML) method was proposed to estimate the covariance of the channel matrix. However, this scheme suffers from low SNR and the BA is achieved only at the UE side because of isotropic probing 
at the BS. In our previous work \cite{sxsBA2017}, we proposed an efficient BA scheme that  jointly estimates the two-sided AoA-AoD of the strongest path from the second-order statistic of the channel matrix. A limitation of \cite{sxsBA2017} as well as most works
based on OFDM signaling \cite{Rappaport2017lowrank,RobertSOMP2017} is the assumption of perfect OFDM frame synchronization and no inter-carrier interference. This is in fact difficult to achieve at mmWaves due to the potentially large multipath delay spread, 
Doppler shifts, and very low SNR  before BA.
These weaknesses, together with the fact that OFDM signaling suffers from large {\em Peak-to-Average Power Ratio} (PAPR), has motivated the proposal of
single-carrier transmission \cite{Ghosh2014singleCarrier,Colavolpe2017singleCarrier} as a more favorable option at mmWaves.
Recently, \cite{AlkhateebTimeDomain2017,Time2017} proposed a time-domain BA approach based on CS techniques for single-carrier mmWave systems.
However, as in \cite{AhmedFreqOMP2015, RobertSOMP2017}, this work focuses on  estimating the instantaneous complex channel coefficients,
with the assumption that these complex coefficients remain invariant over the whole training stage, which is an unrealistic assumption, as discussed above \cite{WeilerMeasure2014,HeathVariation2017,Rappaport2017lowrank, sxsBA2017}.

\subsection{Contributions}

In this paper, we propose an efficient BA scheme for single-carrier mmWave communications with HDA transceivers 
and frequency-selective multipath channels. 
In the proposed scheme, each UE  independently estimates its best AoA-AoD pair over the reserved beacon slots (see Section \ref{systemmodel}), 
during which the BS periodically broadcasts its probing time-domain sequences. We exploit the sparsity of the mmWave channel in both angle and delay domains \cite{NasserAngleTime2016} to reduce the training overhead. We also pose the estimation of the strongest AoA-AoD pair as a {\em Non-Negative Least Squares} (NNLS) problem, which can be efficiently solved by standard techniques. Our main contribution can be summarized as follows:

{1) {\em Pure Time-Domain Operation.} Unlike our prior work in \cite{sxsBA2017} and other works based on OFDM signaling \cite{Rappaport2017lowrank,RobertSOMP2017}, the scheme proposed in this paper takes place completely in the 
time-domain and uses \textit{Pseudo-Noise} (PN) sequences with \textit{good} correlation properties that suits single-carrier mmWave systems. 

{2) {\em More General and Realistic mmWave Channel Model.} We consider a quite general  mmWave wireless channel model, taking into account 
the fundamental features of mmWave channels such as fast time-variation due to Doppler, frequency-selectivity, and the AoA-AoD sparsity \cite{WeilerMeasure2014,Rappaport2017lowrank,Tim2018}. As in \cite{sxsBA2017, Rappaport2017lowrank}, we design a suitable signaling scheme to collect quadratic measurements,  yielding estimates of the channel second-order statistics in the AoA-AoD domain. Consequently, the proposed scheme is highly robust to the channel time-variations. 

{3) {\em Tolerance to Large Doppler shifts.} Unlike our prior work in \cite{sxs2017Time} and the work in \cite{OverviewHeath2016}, which model Doppler as a piecewise constant phase shift changing across blocks of symbols, here we consider a continuous linear (in time) phase shift within the whole beacon slot. 
As a by-product of our refined Doppler model, we notice that longer PN sequences do not necessarily exhibit better performances since they undergo a larger  phase rotations	due to the Doppler. We illustrate by numerical experiments that there is an optimal sequence length based on the given set of parameters, using which the proposed scheme achieves better performances in the presence of large Doppler shifts encountered at mmWaves.
	
{4) {\em Effectiveness of Single-Carrier Modulation.} Our proposed time-domain BA scheme is tailored to single-carrier mmWave systems. In particular, we show that, after achieving BA, the effective channel reduces essentially to a single path with a single delay and Doppler shift, with relatively large SNR due to the high beamforming gain. This means that single-carrier modulation needs no time-domain equalization and the baseband signal processing
becomes very simple, since it requires only standard timing and {\em carrier frequency offset} (CFO) estimation and compensation. 

{\bf Notation}: We  denote vectors by boldface small (e.g., $\bfa$) and matrices by boldface capital (e.g., $\bfA$) letters.  Scalars are denoted by non-boldface letters (e.g., $a$, $A$). We represent sets by calligraphic letter $\clA$ and their cardinality with $|\clA|$. We use $\bE$ for the expectation, $\otimes$ for the  Kronecker product of two matrices, $\bfA^\transp$ for transpose, $\bfA^*$ for conjugate, and $\bfA^\herm$ for conjugate transpose 
of a matrix $\bfA$.  For an integer $k\in\intgr$, we use the shorthand notation $[k]$ for the set of non-negative integers $\{1,...,k\}$.

\section{Problem Statement}

In this section, we provide a general overview of the BA problem based on the channel second-order statistics.
\begin{figure}[t]
	\centerline{\includegraphics[width=8.5cm]{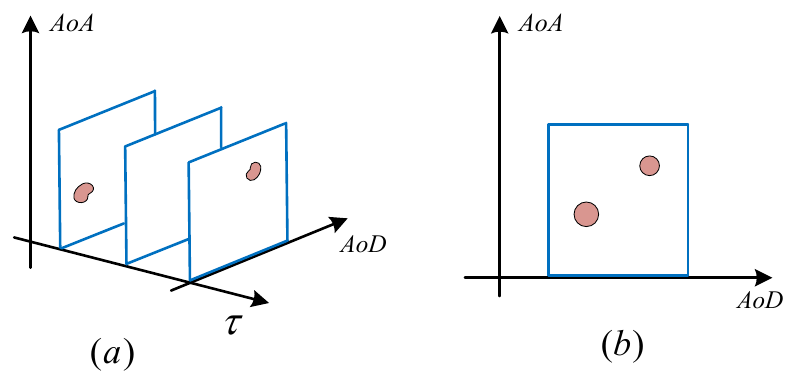}}
	\caption{{\small  Illustration of the channel sparsity in the {\em Angle of Arrival} (AoA), {\em Angle of Departure} (AoD), and delay ($\tau$) domains. (a) Slices of the channel power spread function over discrete delay taps, where only a few slices contain scattering components with large power. (b) Marginal power spread function of the channel in the AoA-AoD domain obtained from the integration of the power spread function along all the delay taps.}}
	\label{1_bread}
\end{figure}
\subsection{Channel Second-Order Statistics}
We consider a standard and widely used mmWave scattering channel (e.g., see \cite{Rappaport2014Capacity,Sayeed2014Sparse}) 
illustrated in \figref{1_bread} (a).
The propagation channel  between the BS and a generic UE consists of  a sparse collection of multipath
components in the AoA-AoD-delay $(\phi,\theta,\tau)$ domain, including a possible {\em Line-of-Sight} (LOS) component as well as
some {Non-Line-of-Sight} (NLOS) reflected paths \cite{Colavolpe2017singleCarrier}.
The scattering channel is modeled as locally {\em Wide-Sense Stationary} (WSS)
process with \textit{Power Spread Function} (PSF) $f_p(\phi, \theta, \tau)$ at the AoA-AoD-Delay $[\phi, \phi + d\phi) \times [\theta, \theta + d\theta) \times [\tau, \tau + d\tau)$.
The PSF encodes the second-order statistics of the channel and it is locally time-invariant as long as the propagation geometry does not change significantly.
The time scale over which the PSF is time-invariant is very large with respect to the inverse of the signaling bandwidth, justifying  the locally WSS assumption.
Practical channel measurements have shown that only a few tapped delay elements are enough to represent the sparse channel \cite{Rappaport2014Capacity,Sayeed2014Sparse,NasserAngleTime2016}. This is illustrated in \figref{1_bread} (a), where only a few slices of the PSF contain
scattering components with large power. The marginal PSF of the channel in the AoA-AoD domain
is obtained by integrating over the whole delay taps as
\begin{align}
f_p( \phi, \theta)=\int_{\tau} f_p(\phi, \theta, d\tau),\label{gam_th_th}
\end{align}
and it is typically very sparse (see, e.g., Fig.\,\ref{1_bread} (b)).

\subsection{Beam-Alignment Using Second-order Statistics}
In terms of BA, we are interested in finding an AoA-AoD pair corresponding to
strong communication path between the UE and the BS.
If the marginal PSF of the channel in the AoA-AoD domain  $f_p(\phi, \theta)$ as in \eqref{gam_th_th} is {\em a-priori} known, the BA problem simply boils down to finding the support of $f_p(\phi, \theta)$
(e.g., see the two bubbles in \figref{1_bread} (b)). In practice, however, $f_p(\phi, \theta)$ is not a priori known and should be
estimated via a suitable signaling scheme. With this in mind, we can pose the BA problem as follows.

\vspace{1mm}
\noindent
\colorbox{gray!40}{
	\begin{minipage}{1.0\textwidth}
		\noindent{\bf BA Problem}: Design a suitable signaling between the BS and the UE, find an estimate of the AoA-AoD PSF $f_p(\phi, \theta)$, and identify an AoA-AoD pair $(\phi_0, \theta_0)$ with a sufficiently large strength $f_p(\phi_0,\theta_0)$. 	
	\end{minipage}
}
\vspace{1mm}

In this paper, we use pseudo-random waveforms with nice auto- / cross-correlation properties as the channel probing signal. We will show that, using the proposed signaling,
each UE is able to collect its own quadratic measurements which yields a noisy linear projection of the PSF in the AoA-AoD domain $f_p(\phi, \theta)$.
We exploit the sparsity and non-negativity of the PSF to reformulate the estimation of $f_p(\phi, \theta)$ as a NNLS problem, which yields a good estimate of the channel second-order statistics.

\begin{figure*}[t]
	\centerline{\includegraphics[width=14cm]{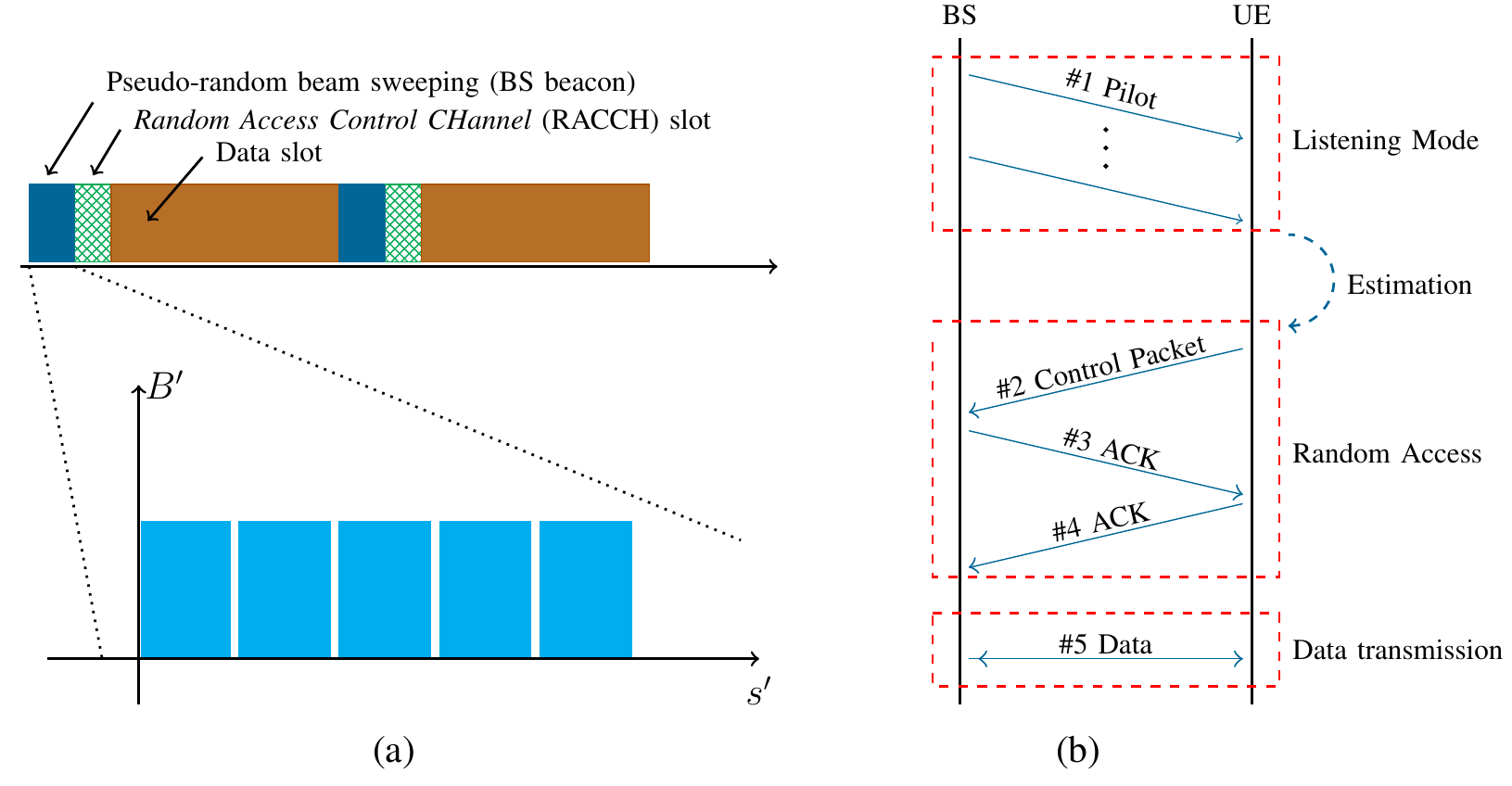}}
\caption{{\small (a) (Top) Frame structure of the proposed \textit{Beam Alignment} (BA) scheme. (Bottom) Each beacon slot consists of $S$ PN sequences indexed by $s'\in [S]$, and all the PN sequences have access to the whole effective bandwidth $B'\leq B$. (b) Illustration of the proposed BA process between the BS and a generic UE, where the procedures (\#2$\sim$\#5) are independently done at each UE, and all the UEs share the same BS beamforming codebook (\#1).}}
\label{2_frame_timeline}
\end{figure*}

\figref{2_frame_timeline} (a) illustrates the proposed frame structure which consists of three parts: the downlink beacon slot, the {\em Random Access Control CHannel} (RACCH) slot, and the data slot. An overview of the proposed initial acquisition and BA protocol is illustrated in \figref{2_frame_timeline} (b). As in \cite{sxsBA2017,sxs2017Time}, the measurements are collected at the UEs from downlink beacon slots broadcasted by the BS.	Each UE selects its strongest AoA-AoD pair $(\phi_0, \theta_0)$ in the estimated $f_p(\phi, \theta)$ as the joint beamforming directions during the data transmission. Then, the  acquisition protocol proceeds as described in  \cite{sxsBA2017,sxs2017Time}.	Namely, the UE sends a beamformed packet to the BS in the RACCH slot, during which the BS stays in listening mode. This packet contains basic information such as user ID and the index of the beam corresponding to the selected AoD $\theta_0$. The BS responds with a data packet with piggybacked acknowledgment in the data subslot of a next frame, and from this moment on the BS and the UE are connected. Further beam refinement and tracking is possible to adapt to small variations of the propagation geometry. However, this can be achieved by rather standard array processing and goes outside the scope of this paper.

\subsection{Equivalent Channel after Beam-Alignment}
After completing a BA cycle as described above, the UE and the BS focus their beams
on a specific AoA-AoD pair $(\phi_0, \theta_0)$ to boost the SNR as much as possible.
As a result, the equivalent channel after beamforming along $(\phi_0, \theta_0)$ can be represented by a \textit{Single-Input Single-Output} (SISO) channel.
The PSF of the resulting SISO channel can be well approximated by $f_p(\tau)=f_p(\phi_0, \theta_0, \tau)$ in the delay domain, which simply corresponds to the PSF obtained
by focusing the transmitted signal power along the estimated AoA-AoD $(\phi_0, \theta_0)$. Due to the underlying channel sparsity \cite{Rappaport2014Capacity,Sayeed2014Sparse,NasserAngleTime2016}, we expect that, after BA, the PSF $f_p(\tau)$ consists of almost a single scattering element with a specific delay $\tau_0$
and Doppler shift $\nu_0$, which can be estimated and compensated by a standard timing and frequency offset synchronization subsystem. Since this is operated after BA, the operating SNR is not at all critical. Therefore, standard techniques for single-carrier synchronization can be used. Furthermore, since the effective channel	after BA reduces to a single multipath component, it is essentially frequency-flat. It follows that near-optimal performance can be achieved by single-carrier communication without the need of equalization. This is confirmed by the results in Section \ref{results}, where we use the effective SISO channel to derive upper and lower bounds on the achievable ergodic rate after BA, showing that time-domain equalization is effectively not needed.


\section{Mathematical Modeling}\label{systemmodel}

\subsection{Channel Model}
Consider a generic UE in a mmWave system served by a specific BS. Suppose that the BS is equipped with a {\em Uniform Linear Array} (ULA) having $M$ antennas and $M_{\text{RF}}\ll M$ RF chains. The UE also has a ULA  with $N$ antennas and $N_{\text{RF}}\ll N$ RF chains.
We assume that both the BS and the UE apply a hybrid beamforming consisting of an analog precoder/combiner and a digital precoder/combiner. In this paper, we will focus mainly on training the  analog precoders/combiners in the initial BA phase. We assume that the propagation channel between the BS and the UE consists of $L\ll \max\{M,N\}$ multipath components,
where the $N\times M$ baseband equivalent impulse response of the channel at time slot $s$ is given by
\begin{align}\label{ch_mod_disc_mp}
\sfH_s(t,\tau)&=\sum_{l=1}^L \rho_{s,l} e^{j2\pi \nu_{l}t}\bfa_{\text{R}}(\phi_l) \bfa_{\text{T}}(\theta_l)^\herm \delta(\tau-\tau_l),
\end{align}
where $(\phi_l, \theta_l, \tau_l, \nu_l)$ denote the AoA, AoD, delay, and Doppler shift 
of the $l$-th component, and $\delta(\cdot)$ denotes the Dirac delta function.
The vectors $ \bfa_{\text{T}}(\theta_l)\in \bC^{M}$  and $ \bfa_{\text{R}}(\phi_l)\in \bC^{N}$ are the array response vectors of the BS and UE at AoD $\theta_l$ and AoA $\phi_l$ respectively, with elements given by
\begin{subequations}  \label{array-resp}
	\begin{align}
	[\bfa_{\text{T}}(\theta)]_m&=e^{j (m-1)\pi \sin(\theta)}, m \in[M], \label{a_resp_BS}\\
	[\bfa_{\text{R}}(\phi)]_n&=e^{j (n-1)\pi \sin(\phi)},\  n\in[N],\label{a_resp_UE}
	\end{align}
\end{subequations}
where we assume that the spacing of the ULA antennas equals to half wavelength.

For simplicity, in the channel model \eqref{ch_mod_disc_mp}, we neglect the effect of pulse shaping 
and assume a frequency-flat pulse response over the signal bandwidth \cite{Rappaport2017lowrank}.
Also, for the sake of modeling simplicity, we assume here that each multipath component has a very narrow footprint over the AoA-AoD and delay domain.
Extension to more widely spread multipath clusters is straightforward and will be applied in the numerical simulations.
Moreover, we make the very standard assumption in array processing that the array response vectors are invariant with 
frequency over the signal bandwidth (i.e., that wavelength $\lambda$ over the frequency interval $f\in[f_0-B/2,f_0+B/2]$ 
can be approximated as $\lambda_0 = c/f_0$ where $c$ denotes the speed of light. 
This is indeed well verified when $B$ is less than $1/10$ of the carrier frequency 
(e.g., $B = 1$GHz with carrier between $30$ and $70$ GHz).
Each scatterer corresponding to a AoA-AoD-Delay $(\phi_l, \theta_l, \tau_l)$ has a Doppler shift $\nu_l = \frac{\Delta v_{l}f_0}{c}$ where
$\Delta v_{l}$ indicates the relative speed of the receiver, the $l$-th scatterer, and the transmitter \cite{OverviewHeath2016}. 
We adopt a block fading model, where the channel gains $\rho_{s,l}$ remain invariant over the
channel \textit{coherence time} $\Delta t_c$ but change i.i.d. randomly across different \textit{coherence times} \cite{Rappaport2017lowrank}.
Since each scatterer in practice is a superposition of many smaller components that have (roughly) the same AoA-AoD and delay, we assume a general 
Rice fading model given by
\begin{align}\label{rice_fading}
\rho_{s,l}\sim \sqrt{\gamma_l} \left(\sqrt{\frac{\eta_l}{1+\eta_l}}+\frac{1}{\sqrt{1+\eta_l}}\check{\rho}_{s,l}\right),
\end{align}
where $\gamma_l$ denotes the overall multipath component strength, 
$\eta_l\in[0,\infty)$ indicates the strength ratio between the LOS and the NLOS components, 
and $\check{\rho}_{s,l} \sim \cg(0, 1)$ is a zero-mean unit-variance complex Gaussian random variable. 
In particular, $\eta_l\to \infty$ indicates a pure LOS path while  $\eta_l=0$ indicates a pure NLOS path, affected by standard Rayleigh fading.

\subsection{Proposed Signaling Scheme}\label{signal_procedure}

We assume that the BS can simultaneously transmit up to $M_{\text{RF}}\ll M$ different pilot streams.
In this paper, we consider a time-domain signaling where a unique PN sequence is assigned to each RF chain (pilot stream),  similar to standard {\em Code Division Multiple Access} (CDMA).
Unlike our previous work in \cite{sxsBA2017}, where different pilot streams are assigned with sets of orthogonal subcarriers, such that
(in the absence of inter-carrier interference) they can be perfectly separated by the UE in the frequency domain,
in the proposed scheme, different pilot streams are generally non-orthogonal (non-separable) but become almost separable
if the assigned PN sequences have good cross-correlation properties and are sufficiently long.
Let $x_{s,i}(t)$, $t \in [st_0, (s+1)t_0)$, be the continuous-time baseband equivalent PN signal corresponding
to the $i$-th ($i\in[M_{\text{RF}}]$) pilot stream transmitted over $s$-th slot, given by
\begin{align}\label{time_PN_sig}
x_{s,i}(t)=\sum_{n=1}^{N_c}\varrho_{n,i}p_{r}(t-nT_c),\quad \varrho_{n,i}\in\{1,-1\},
\end{align}
where $t_0$ denotes the duration of the PN sequence, $p_{r}(t)$ is the normalized band-limited pulse shaping filter response with normalized energy $\int |p_{r}(t)|^2dt=1$. We assume that the PN sequence has a chip duration of $T_c$, a bandwidth of $B'=1/T_c \leq B$, and a total of $N_c=t_0/T_c=t_0B'$ chips, where $B$ is the maximum available bandwidth. We shall choose a suitable PN sequence length $N_c$, such that the resulting time-domain signal \eqref{time_PN_sig} is transmitted in a sufficiently small time-interval $t_0$ over which the channel can be considered time-invariant, i.e., $t_0\leq \Delta t_c$.

To transmit the $i$-th pilot stream, the BS applies a beamforming vector $\bfu_{s,i} \in \bC^M$. Without loss of generality, the beamforming vectors are normalized such that $\|\bfu_{s,i}\|=1$. As mentioned before, we consider a HDA beamforming architecture where the beamforming function is implemented in the analog RF domain. Hence,
the beamforming vectors $\bfu_{s,i}$, $i \in  [M_{\text{RF}}]$, are independent of frequency and constant over the whole bandwidth.
The transmitted signal at slot $s$ is given by
\begin{align}
\bfx_s(t) & =\sum_{i=1}^{M_{\text{RF}}}\sqrt{\frac{\ptot T_c}{M_{\text{RF}}}} x_{s,i}(t) \bfu_{s,i}
= \sum_{i=1}^{M_{\text{RF}}}\sum_{n=1}^{N_c}\sqrt{\frac{\ptot T_c}{M_{\text{RF}}}}\varrho_{n,i}p_{r}(t-nT_c)\bfu_{s,i},
\end{align}
where $\ptot$ is the total transmit power which is equally distributed into the $M_{\text{RF}}$ RF chains from BS. Consequently, the received basedband equivalent signal at the UE array is
\begin{align}\label{receiveTT}
\bfr_s(t)&=\int \sfH_s(t,d\tau) \bfx_s(t-\tau) =\sum_{l=1}^L \sfH_{s,l}(t)\bfx_s(t-\tau_l)\nonumber\\
&=\sum_{i=1}^{M_{\text{RF}}} \sum_{l=1}^L\sqrt{\frac{\ptot T_c}{M_{\text{RF}}}}  \sfH_{s,l}(t)x_{s,i}(t-\tau_l)\bfu_{s,i} ,
\end{align}
where $\sfH_{s,l}(t) := \rho_{s,l} e^{j2\pi \nu_{l}t} \bfa_{\text{R}}(\phi_l) \bfa_{\text{T}}(\theta_l)^\herm$, $l\in[L]$ are the time-varying
MIMO channel ``taps'' corresponding to the $L$ multipath components \eqref{ch_mod_disc_mp}.

With a hybrid MIMO structure, the UE does not have direct access to (a sampled version of) the components of $\bfr_s(t)$. Instead, at each measurement slot $s$, the UE must apply some beamforming vector in the analog domain obtaining a projection
of the received signal. Since the UE has  $N_{\text{RF}}$ RF chains, it can obtain up to $N_{\text{RF}}$ such projections per slot.
The analog RF signal received at the UE antenna array is distributed across the $N_{\text{RF}}$ RF chains for demodulation.
This is achieved by signal splitters that divide the signal power by a factor of $N_{\text{RF}}$.
Thus, the received signal at the output of the $j$-th RF chain at the UE side is given by
\begin{align}\label{eq:j_out}
y_{s,j}(t)&=\!\frac{1}{\sqrt{N_{\text{RF}}}} \bfv_{s,j}^\herm \bfr_s(t)+z_{s,j}(t) =\!\!\sum_{i=1}^{M_{\text{RF}}}\sum_{l=1}^L\!\!\sqrt{\pdim}\bfv_{s,j}^\herm\sfH_{s,l}(t)x_{s,i}(t\!-\!\tau_l)\bfu_{s,i}\!+\!z_{s,j}(t),
\end{align}
where $\pdim = \frac{\ptot T_c}{M_{\text{RF}}N_{\text{RF}}}$ indicates the power distribution to the multiple RF chains on both sides, ${\bfv_{s,j} \in \bC^N}$ denotes the normalized beamforming vector of the $j$-th RF chain at the UE side with $\|\bfv_{s,j}\|=1$, $z_{s,j}(t)$ is the continuous-time complex {\em Additive White Gaussian Noise} (AWGN) at the output of the $j$-th RF chain, with a {\em Power Spectral Density} (PSD) of $N_0$ Watt/Hz. The noise at the receiver is mainly introduced by the RF chain electronics, e.g., filter, mixer, and A/D conversion. The factor $\frac{1}{\sqrt{N_{\text{RF}}}}$ in \eqref{eq:j_out} takes into account the power split said above, assuming that this only applies to the useful signal and not the thermal noise. Therefore, this received signal model is a conservative worst-case assumption.

We adopt a simplified continuous linear Doppler phase shift model in $\sfH_{s,l}(t)$, given by
\begin{align}\label{doppler_model}
\sfH_{s,l}(t)|_{t\in [nT_c,(n+1)T_c)}\approx\rho_{s,l} e^{j2\pi (\check{\nu}_{s,l}+\nu_{l}nT_c)} \bfa_{\text{R}}(\phi_l) \bfa_{\text{T}}(\theta_l)^\herm=\sfH_{s,l}e^{j2\pi \nu_{l}nT_c},\quad n\in[N_c]
\end{align}
where $\sfH_{s,l}:=\rho_{s,l} e^{j2\pi\check{\nu}_{s,l}} \bfa_{\text{R}}(\phi_l) \bfa_{\text{T}}(\theta_l)^\herm$, and where $\check{\nu}_{s,l}$ represents an arbitrary initial value which changes i.i.d randomly over different beacon slots.
As a result, the product term $\sfH_{s,l}(t)x_{s,i}(t-\tau_l)$ in \eqref{eq:j_out} can be written as
\begin{align}\label{x_l}
\sfH_{s,l}(t)x_{s,i}(t\!-\!\tau_l)&=\sfH_{s,l}\!\sum_{n=1}^{N_c}\varrho_{n,i}p_{r}(t\!-\!nT_c\!-\!\tau_l)e^{j2\pi \nu_{l}nT_c}:=\sfH_{s,l} x_{s,i}^l(t-\tau_l),
\end{align}
where $x_{s,i}^l(t)$  is given by
\begin{align}\label{x_l_rotate}
x_{s,i}^l(t)=\sum_{n=1}^{N_c}\varrho_{n,i}p_{r}(t-nT_c)e^{j2\pi \nu_{l}nT_c}.
\end{align}
We can interpret $x_{s,i}^l(t)$ as a rotated version of the original transmitted PN sequence $x_{s,i}(t)$, where the $n$-th chip of the original signal $x_{s,i}(t)$ is rotated by a small Doppler shift $e^{j2\pi \nu_{l}nT_c}$.
Substituting  \eqref{x_l} into \eqref{eq:j_out}, we can write the received signal $y_{s,j}(t)$ in \eqref{eq:j_out}  as
\begin{align}\label{eq:j_out_simp}
y_{s,j}(t)\!=\!\sum_{i=1}^{M_{\text{RF}}}\sum_{l=1}^L\!\!\sqrt{\pdim}\bfv_{s,j}^\herm\sfH_{s,l} x_{s,i}^l(t\!-\!\tau_l)\bfu_{s,i}\!+\!z_{s,j}(t).
\end{align}

Since the PN sequences assigned to the $M_{\text{RF}}$ RF chains are mutually (roughly) orthogonal, the $M_{\text{RF}}$ pilot streams transmitted from the BS side can be approximately separated at the UE by correlating the received signal with a desired matched filter $x_{s,i}^*(-t)=\sum_{n=1}^{N_c}\varrho_{n,i}p_{r}^*(-t+nT_c)$. Consequently, the $i$-th BS pilot stream received through the $j$-th RF chain at the UE is given by
\begin{align}\label{eq:j_outFrom_i}
y_{s,i,j}(t)&=\int y_{s,j}(\tau)x_{s,i}^*(\tau-t)d\tau=\sum_{l=1}^L\sum_{i'=1}^{M_{\text{RF}}}\!\!\sqrt{\pdim}\bfv_{s,j}^\herm\sfH_{s,l} R_{i',i}^{x^l}(t\!-\!\tau_l)\bfu_{s,i}\!+\!z_{s,j}^c(t)\nonumber\\
&\overset{(a)}{\approx} \sum_{l=1}^L\sqrt{\pdim}\bfv_{s,j}^\herm\sfH_{s,l} R_{i,i}^{x^l}(t\!-\!\tau_l)\bfu_{s,i}\!+\!z_{s,j}^c(t)
\end{align} 
where $\forall i,i'\in[M_{\text{RF}}]$, $R_{i',i}^{x^l}(t):=\int x_{s,i'}^l(\tau)x_{s,i}^*(\tau-t)d\tau$ represents the correlation between the Doppler-rotated sequence $x_{s,i'}^l(t)$ given by \eqref{x_l_rotate} and the desired matched filter $x_{s,i}^*(-t)$, and $z_{s,j}^c(t) = \int z_{s,j}(\tau)x_{s,i}^*(\tau-t)d\tau$  denotes the noise at the output of the matched filter. The approximation $(a)$ in \eqref{eq:j_outFrom_i} follows the fact that, the cross-correlations between different PN sequences are nearly zero, i.e., $R_{i'\!,i}^x(t)=\int x_{s,i'}(\tau)x_{s,i}^*(\tau-t)d\tau\approx 0$, for $i'\neq i$. Also note that, the phase rotation for each chip in $x_{s,i}^l(t)$ is very small ($\nu_{l}T_c\ll 1$), hence, we can apply the following approximation $R_{i'\!,i}^{x^l}(t)=\int x_{s,i'}^l(\tau)x_{s,i}^*(\tau-t)d\tau\approx 0$, for $i'\neq i$. In numerical simulations, we will consider the general case where the sequences are not necessarily orthogonal and the Doppler shift can be moderately large.

Consider \eqref{eq:j_outFrom_i} and suppose that the output signal at the UE side is sampled at chip-rate, the resulting discrete-time signal can be written as
\begin{align}\label{eq:j_outFrom_i_sample}
y_{s,i,j}[k]&=y_{s,i,j}(t)|_{t=kT_c}=\!\sum_{l=1}^L\!\!\sqrt{\pdim}\bfv_{s,j}^\herm\sfH_{s,l} R_{i,i}^{x^l}(k T_c\!-\!\tau_l)\bfu_{s,i}\!+\!z_{s,j}^c[k],
\end{align}
where $k\in[\check{N}_c]$, $\check{N}_c\geq N_c+\frac{\Delta \tau_{\max}}{T_c}$ indicates the sampling indices, and where $\Delta \tau_{\max}=\max\{|\tau_l-\tau_{l'}|: {l,l'\in[L]}\}$ denotes the maximum delay spread of the channel.
Note that for PN sequences, the sequence $k \mapsto |R_{i,i}^{x^l}(kT_c\!-\!\tau_l)|$ in \eqref{eq:j_outFrom_i_sample}, seen as a function of the discrete index $k$, has sharp peaks at indices $k_l \approx \frac{\tau_l}{T_c}$ corresponding to the delay of the scatterers. Intuitively speaking, the output $y_{s,i,j}[k]$ at those indices $k_l$ yields Gaussian variables whose power is obtained by projecting the AoA-AoD-Delay PSF $f_p(\phi, \theta, \tau)$ along beamforming vectors $(\bfu_{s,i},\bfv_{s,j})$ in the angular domain and along the $k_l$-th slice in the delay domain with $\tau \in [k_lT_c, (k_l+1)T_c]$. The slicing in the delay domain results from the fact that $|R_{i,i}^{x^l}(kT_c\!-\!\tau_l)|$ is well localized around $k_l$. We refer to \figref{1_bread} (a) for an illustration and will use this property later on in the paper to design our BA algorithm.

\subsection{Sparse Beamspace Representation}

In practice, the AoA-AoDs $(\phi_l, \theta_l)$ in \eqref{ch_mod_disc_mp} can take on arbitrary values in the contnuum of AoA-AoDs. Following the widely used approach of \cite{SayeedVirtualBeam2002}, known as {\em beamspace representation}, we obtain a finite-dimensional representation of the channel by quantizing the antenna-domain channel response \eqref{ch_mod_disc_mp} with respect to a discrete dictionary in the AoA-AoD (angle) domain. More specifically, we consider the discrete set of AoA-AoDs
\begin{subequations} \label{theta-phi}
	\begin{align}\label{gridtheta}
	\Phi&:=\{\check{\phi}: (1+\sin(\check{\phi}))/2=\frac{n-1}{N}, \, n \in [N]\},\\
	\Theta&:=\{\check{\theta}: (1+\sin(\check{\theta}))/2=\frac{m-1}{M}, m\in [M]\},
	\end{align}
\end{subequations}
and use the corresponding array responses  $\clA_{\text{R}}:=\{\bfa_{\text{R}}(\check{\phi}): \check{\phi} \in \Phi\}$ and $\clA_{\text{T}}:=\{\bfa_{\text{T}}(\check{\theta}): \check{\theta} \in \Theta\}$ as a discrete dictionary to represent the channel response. For the ULAs considered in this paper, the dictionary $\clA_{\text{R}}$ and $\clA_{\text{T}}$, after suitable normalization, yields the orthonormal bases that corresponds to the {\em Discrete Fourier Transformation} (DFT) matrices $\bfF_{N}\in \bC^{N\times N}$ and $\bfF_{M}\in \bC^{M\times M}$, respectively, with elements
\begin{subequations}
	\begin{align}
	[\bfF_{N}]_{n,n'}&=\frac{1}{\sqrt{N}}e^{j2\pi (n-1)(\frac{n'-1}{N}-\frac{1}{2})}, n,n'\in[N],\\
	[\bfF_{M}]_{m,m'}&=\frac{1}{\sqrt{M}}e^{j2\pi (m-1)(\frac{m'-1}{M}-\frac{1}{2})}, m,m'\in[M].
	\end{align}
\end{subequations}
Using this discrete quantized dictionaries, we obtain the virtual angle-domain channel representation of $\check{\sfH}_s(t,\tau)$, given by
\begin{align}\label{beamspacechannel}
\check{\sfH}_s(t,\tau)& = \bfF_{N}^\herm\sfH_s(t,\tau)\bfF_{M} = \sum_{l=1}^{L}\check{\sfH}_{s,l}(t)\delta(\tau-\tau_l),
\end{align}
where $\check{\sfH}_{s,l}(t) := \bfF_{N}^\herm\sfH_{s,l}(t)\bfF_{M}$. We have shown in our earlier work \cite{sxsBA2017} that, as the number of antennas $M$ at the BS and $N$ at the UE increases, the DFT basis provides a good sparsification of the propagation channel. As a result, $\check{\sfH}_s(t,\tau)$ can be approximated as a $L$-sparse matrix, with $L$ non-zero elements in the locations corresponding to the AoA-AoDs of the $L$ scatterers. We may encounter a grid error in \eqref{beamspacechannel}, since the AoAs/AoDs do not necessarily fall into the uniform grid $\Phi\times\Theta$. Nevertheless, as shown in \cite{sxsBA2017}, the grid error becomes negligible by increasing the number of antennas (grid resolution). We will evaluate this off-grid effect in the numerical results.

\section{Proposed Beam Alignment Scheme}\label{proposedscheme}

\subsection{BS Channel Probing and UE Sensing}

Consider the scattering channel model in \eqref{ch_mod_disc_mp} and its virtual angle-domain representation in \eqref{beamspacechannel}. In our proposed scheme, at each beacon slot $s$, the BS  probes the channel along $M_{\text{RF}}$ beamforming vectors $\bfu_{s,i}$, $i\in[M_{\text{RF}}]$, each of which is applied to a unique PN sequence signal $x_{s,i}(t)$. We select the beamforming vectors at the BS side according to a pre-defined pseudo-random codebook, which is a collection of the angle sets $\clC_\text{T}:= \{\Uc_{s,i} : s \in [T], i \in [M_{\text{RF}}] \}$, where $\Uc_{s,i}$ denotes the angle-domain support of the beamforming vector $\bfu_{s,i}$, i.e., the indices of the quantized angles in the virtual angle-domain representation of $\bfu_{s,i}$, and where $T$ is the effective period of beam training. We assume that the beamforming vector $\bfu_{s,i}$ sends equal power along the directions in $\Uc_{s,i}$ with the number of active angles given by $|\Uc_{s,i}|=:\kappa_u\leq M$, which we assume to be the same for all  $(s,i)$. We call $\kappa_u$ the power spreading factor or the transmit ``beamwidth''.  Consequently, we obtain the beamforming vectors at the BS given by  $\bfu_{s,i} = \Fm_M \check{\bfu}_{s,i}$, where
$\check{\bfu}_{s,i}=\frac{\one_{\clU_{s,i}}}{\sqrt{\kappa_u}}$, and where $\one_{\clU_{s,i}}$ denotes a vector with $1$
at components in the support set  $\clU_{s,i}$ and $0$ elsewhere. One can simply imagine the vector $\check{\bfu}_{s,i}$ as a finger-shaped beam pattern  in the angle-domain as illustrated in \figref{cluster} (a). We assume that the angle indices in $\Uc_{s,i}$ in the codebook $\clC_\text{T}$ are a priori generated in a random manner and shared to all the UEs in the system, thus, we call $\clC_\text{T}$ a pseudo-random codebook.
\begin{figure*}[t]
	\centering
	\includegraphics[width=14cm]{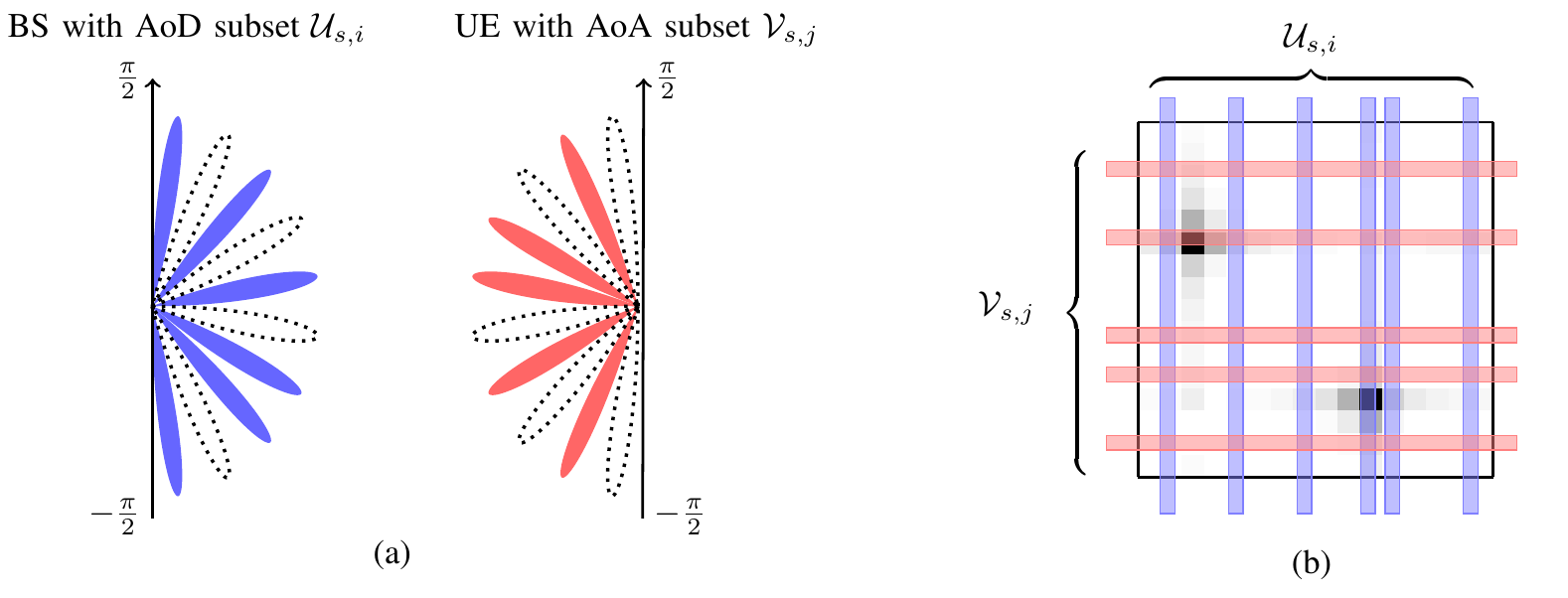}
	\caption{{\small {\em (a) Illustration of the subset of AoA-AoDs at time slot $s$ probed by the $i$-th beacon stream
				transmitted by the BS and received by the $j$-th RF chain of the UE, for $M = N = 10$. 
				The AoD subset is given by  $\clU_{s,i}=\{1,3,4,6,8,10\}$ { (numbered counterclockwise)} with beamforming vector 
				$\check{\bfu}_{s,i}=\frac{1}{\sqrt{6}}[1,0,1,0,1,0,1,1,0,1]^\transp$. The AoA subset is given by 
				$\clV_{s,j}=\{2,4,5,7,9\}$ { (numbered counterclockwise)} with receive beamforming vector $\check{\bfv}_{s,j}=\frac{1}
				{\sqrt{5}}[0,1,0,1,1,0,1,0,1,0]^\transp$.
				(b) The channel gain matrix $\check{\Gammam}$ (with two strong MPCs indicated by the dark spots) 
				measuring along $\clV_{s,j} \times \clU_{s,i}$}}.}
	\label{cluster}
\end{figure*}
At the UE side, each UE can locally customize its own receive beamforming codebook defined as $\clC_\text{R}:= \{ \Vc_{s,j} : s \in [T], j \in [N_{\text{RF}}]\}$, where $\Vc_{s,j}$, with $|\Vc_{s,j}| = \kappa_v \leq N$ for all $(s,j)$, is the angle-domain support, defining the directions from which the receiver beam patterns collect the signal power. We define the beamforming vectors at the UE side by $\bfv_{s,j} = \Fm_N \check{\bfv}_{s,j}$, where $\check{\bfv}_{s,j}=\frac{\one_{\clV_{s,j}}}{\sqrt{\kappa_v}}$ again defines the finger-shaped beam patterns as shown in \figref{cluster} (a). Similar to the power spreading factor $\kappa_u$ at the BS, the parameter $\kappa_v$  controls the spread of the sensing beam patterns at the UE.

Note that in our proposed scheme,  the UEs collect their measurements independently and simultaneously, without any influence or coordination to each other. Therefore, the proposed scheme is quite scalable for multiuser scenarios, where the overhead of training all the UEs does not increase with the number of UEs. This is obviously superior to traditional multi-level BA schemes, where the training overhead grows proportionally with the number of UEs.

\subsection{UE Measurement Sparse Formulation}

During the $s$-th beacon slot, the UE applies the receive beamforming vector $\bfv_{s,j}$ to its $j$-th RF chain. Assuming that the probing PN signals $x_{s,i}(t)$ are approximately orthogonal in the time domain as before, each RF chain at the UE side can almost perfectly separate the transmitted $M_{\text{RF}}$ pilot streams. Thus, using the virtual channel representation in \eqref{beamspacechannel}, we can write \eqref{eq:j_outFrom_i_sample} as 
\begin{align}\label{eq:j_outFrom_i_sample_beamspace}
y_{s,i,j}[k]\!=\!\sum_{l=1}^L\!\!\sqrt{\pdim}\check{\bfv}_{s,j}^\herm\check{\sfH}_{s,l}R_{i,i}^{x^l}(kT_c\!-\!\tau_l)\check{\bfu}_{s,i}\!+\!z_{s,j}^c[k],
\end{align}
where $\check{\bfu}_{s,i}=\bfF_M^\herm \bfu_{s,i}$ and $\check{\bfv}_{s,j}= \bfF_N^\herm {\bfv}_{s,j}$ are the beamforming vectors in the virtual beam domain. Here, we used the unitary property of the DFT matrices, i.e., $\Fm_M^\herm\Fm_M=\bfI_{M}$ and $\Fm_N^\herm\Fm_N=\bfI_{N}$, where $\bfI_{M}$ and $\bfI_{N}$ are identity matrices of dimension $M$ and $N$ respectively.

To formulate the sparse estimation problem, we define   $\check{\bfh}_{s,l} = 1/\sqrt{N_{\text{RF}}}\cdot\vec{(\check{\sfH}_{s,l})}$, $l\in[L]$, as the channel vectors corresponding to the $L$ paths contained in the whole propagation channel and result in a reformulated channel matrix   $\check{\bfH}_s=[\check{\bfh}_{s,1},\,\cdots\, ,\check{\bfh}_{s,L}]$, where $\vec(\cdot)$ denotes the vectorization operator.  We also define a vector $\bfc^{i}_k=[R_{i,i}^{x^1}(kT_c-\tau_1),\,\cdots\,,R_{i,i}^{x^L}(kT_c-\tau_L)]^\transp\cdot \sqrt{\pdim}$, which can be regarded as the \textit{Power Delay Profile} (PDP) of the $i$-th pilot stream transmitted along the $L$ paths and sampled at the $k$-th delay tap $k T_c$.   Consequently, we can express the received beacon signal \eqref{eq:j_outFrom_i_sample_beamspace} at the UE as
\begin{align}\label{eq:j_out_beamspace}
y_{s,i,j}[k]&=\sum_{l=1}^L\!\!\sqrt{\pdim}\check{\bfv}_{s,j}^\herm\check{\sfH}_{s,l}R_{i,i}^{x^l}(kT_c\!-\!\tau_l)\check{\bfu}_{s,i}\!+\!z_{s,j}^c[k]=(\check{\bfu}_{s,i}\otimes \check{\bfv}^*_{s,j})^\transp \check{\bfH}_s \bfc^{i}_k+z_{s,j}^c[k]\nonumber\\
&=\bfg_{s,i,j}^\transp \check{\bfH}_s \bfc^{i}_k+z_{s,j}^c[k],
\end{align}
where we used the well-known identity $\vec(\bfA \bfB \bfC)=(\bfC^\transp \otimes \bfA) \vec(\bfB)$, and where $\bfg_{s,i,j}:=\check{\bfu}_{s,i}\otimes \check{\bfv}^*_{s,j}$ denotes the combined angle-domain beamforming vector corresponding to the $i$-th RF chain at the BS and the $j$-th RF chain at the UE. 

Note that for the high data rates at mmWaves, e.g., the chip rate used in IEEE 802.11ad preamble is $1760$ MHz \cite{Time2017}, it is impractical to use different beamforming vectors for consecutive symbols within the same beacon slot since this would involve a very fast switching of the analog RF beamforming network, which is either impossible or too power consuming to implement.
In a more flexible way, we assume that each beacon slot consists of $S$ symbols, during which the combined beamforming vector $\bfg_{s,i,j}$ remains constant whereas $\check{\bfH}_s$ only changes because of the Doppler shifts $\nu_{l}$. Over different beacon slots, in contrast, we assume that the beamforming vector $\bfg_{s,i,j}$ changes periodically according to the
pre-defined pseudo-random beamforming codebook $\clU_{s,i} \times \clV_{s,j}$. With a slight abuse of notation, we index the received symbols belonging to the $(s+1)$-th beacon slot as $sS+s'$, $s'\in[S]$. It follows that the received signal through the $i$-th RF chain at the BS and the $j$-th RF chain at the UE after matched filtering (refer to \eqref{eq:j_out_beamspace}) can be written as
 \begin{align}\label{eq:j_out_beamspaceT1T2}
 y_{sS+s',i,j}[k]=\bfg_{s,i,j}^\transp \check{\bfH}_{sS+s'} \bfc^{i}_k+z_{sS+s',j}^c[k].
 \end{align} 

To ensure that the proposed BA scheme is highly robust to the rapid channel variations \cite{Rappaport2017lowrank}, we focus on the second-order statistics of the channel coefficients. More specifically, we accumulate the energy at the output of the matched filter across all the $\check{N}_c$ delay taps, given by
\begin{align}\label{q_check}
&\check{q}_{sS+s',i,j}=\sum_{k=1}^{\check{N}_c}|y_{sS+s',i,j}[k]|^2\nonumber\\
&=\bfg_{s,i,j}^\transp\!\left(\!\sum_{l=1}^{L}\check{\bfh}_{sS+s',l}\check{\bfh}_{sS+s',l}^\herm\!\sum_{k=1}^{\check{N}_c}\pdim|R_{i,i}^{x^l}(kT_c\!-\!\tau_l)|^2\!\right)\!\bfg_{s,i,j}\nonumber\\
&+\sum_{k=1}^{\check{N}_c}|z_{sS+s',j}^c[k]|^2+\sum_{k=1}^{\check{N}_c}\xi_{sS+s',i,j}^{h}+\sum_{k=1}^{\check{N}_c}2\xi_{sS+s',i,j}^{z},
\end{align}
where the first two terms represent the signal and the noise contributions respectively. Note that in the signal part in \eqref{q_check}, only $L$ out of $\check{N}_c$ slices which correspond to the $L$ scatterers in the delay domain contains signal power, whereas all the other slices are approximately zero due to the low cross-correlation property of PN sequences. Moreover, the remaining two terms in \eqref{q_check} are given by
\begin{align}\label{crossH}
\xi_{s,i,j}^{h} &=\sum_{l\neq l'}^L \pdim\cdot \bfg_{s,i,j}^\transp\check{\bfh}_{sS+s',l}\check{\bfh}_{sS+s',l'}^\herm  R_{i,i}^{x^l}(kT_c\!-\!\tau_l) R_{i,i}^{x^l}(kT_c\!-\!\tau_{l'})^\herm\bfg_{s,i,j},
\end{align}
\begin{align}\label{crosshZ}
\xi_{sS+s',i,j}^z=2\Re\left \{ \bfg_{s,i,j}^\herm \check{\bfH}_{sS+s'} \bfc^{i}_k \cdot z_{sS+s',j}^c[k]^\herm \right \},
\end{align}
respectively, where \eqref{crossH} denotes the cross product of channel vectors corresponding to different paths, and \eqref{crosshZ} denotes the cross product between channel vectors and the noise.

To obtain a more reliable statistical measurement, we take an average of \eqref{q_check} over the $S$ sequences within each beacon slot. Note that the channel coefficients corresponding to different paths are independent, also, the channel coefficients and the noise are always independent of each other. Consequently, the cross terms \eqref{crossH} and \eqref{crosshZ} have a zero mean. Thus, when the number of symbols $S$ (over which the instantaneous energy $\check{q}_{sS+s',i,j}$ is averaged) is large, theses cross-terms contribute negligibly to \eqref{q_check} and can be treated as a small residual term. As a result, we obtain the following approximation in each beacon slot $s$, given by
\begin{align}\label{q_nocheck}
q_{s,i,j}&=\frac{1}{S}\sum_{s'=1}^{S} \check{q}_{sS+s',i,j}\nonumber\\
&=\frac{\bfg_{s,i,j}^\transp}{S}\!\!\sum_{s'=1}^{S}\!\!\left(\!\sum_{l=1}^{L}\check{\bfh}_{sS+s'\!,l}\check{\bfh}_{sS+s'\!,l}^\herm\!\sum_{k=1}^{\check{N}_c}\!\!\pdim|R_{i,i}^{x^l}(kT_c\!-\!\tau_l)|^2\!\right)\!\!\bfg_{s,i,j}\nonumber\\
&+\frac{1}{S}\sum_{s'=1}^{S}\left(\sum_{k=1}^{\check{N}_c}|z_{sS+s',j}^c[k]|^2\right)+w_{s,i,j},
\end{align}
where $w_{s,i,j}$ represents the residual fluctuation obtained from the averaged cross-terms \eqref{crossH} \eqref{crosshZ}.

As we explained in Section \ref{signal_procedure}, neglecting the effect of the noise, the output $y_{s,i,j}[k]$ is a Gaussian variables whose power is obtained by projecting the AoA-AoD-Delay PSF $f_p(\phi,\theta , \tau)$ along beamforming vectors $\bfu_{s,i}$ and $\bfv_{s,j}$  in the angular domain (due to $\bfg_{s,i,j}:=\check{\bfu}_{s,i}\otimes \check{\bfv}^*_{s,j}$) and along the $k$-th  slice corresponding to $\tau \in [kT_c, (k+1)T_c]$ in the delay domain, where  the slicing in the delay domain results from the fact that the correlation function $|R_{i,i}^{x^l}(t)|$ between the Doppler-rotated sequence $x_{s,i}^l(t)$ given by \eqref{x_l_rotate} and the desired matched filter $x_{s,i}^*(-t)$ is well localized around $t=0$; we refer to \figref{1_bread} (a) for an illustration. Consequently, the summation of the instantaneous powers of $y_{sS+s',i,j}[k]$ along all the delay taps $k \in [N_c]$ yields an estimate of  PSF in the AoA-AoD domain as in \eqref{gam_th_th}.
This has been illustrated in Fig.\,\ref{cluster} (b), where it is seen that each projection computes the summation of those AoA-AoD PSF at the grid points lying at the intersection of probing directions $\clU_{s,i}$ at the BS and $\clV_{s,j}$ at the UE. In the following, we provide a more rigorous formulation of this property.

Without loss of generality, we assume that the energy contained in each PN sequence is constant given by $R^x(0)=R_{i,i}^x(0)=N_c$, $\forall i\in[M_{\text{RF}}]$. Assuming, for simplicity,  that the Doppler phase rotation for each chip is very small ($\nu_{l}T_c\ll 1$), we make the following approximation
\begin{align}\label{R_app}
	|R_{i,i}^{x^l}(t)|&\leq |R_{i,i}^{x^l}(0)|\approx R^x(0), \forall i\in[M_{\text{RF}}].
\end{align}   
where, we assume that, due to the large bandwidth for each chip $B'=1/T_c$, the  matched-filtering loss caused by the Doppler shift is negligible (we take into account all the imperfections due to the Doppler and also non-orthogonal  PN sequences  in the simulations).

Let $\Gammam$ denote the all-zero $N\times M$ matrix with positive elements corresponding to the angle-domain second-order statistics of the channel coefficients, given by
\begin{align}\label{averageGamma}
[\Gammam]_{n,m} = \sum_{l=1}^L\bE\left[|[\check{\sfH}_{sS+s',l}]_{n,m}|^2\right]\cdot \frac{\pdim}{\kappa_u\kappa_v }\cdot |R^{x^l}(0)|^2.
\end{align} 
Also, from the well known property of the autocorrelation function \cite{Gubnerbook}, we have
\begin{align}\label{averageNoise}
\frac{1}{S}\sum_{s'=1}^{S}|z_{sS+s',j}^c[k]|^2\to\bE[|z_{sS+s',j}^c[k]|^2]=N_0R^x(0).
\end{align} 
Consequently, we can approximate \eqref{q_nocheck} by
\begin{align}\label{q_nocheck3}
q_{s,i,j} = \bfb_{s,i,j}^\transp \vec{(\Gammam)} + \check{N}_cN_0R^x(0) + w_{s,i,j},
\end{align}
where $\bfb_{s,i,j} = \bfg_{s,i,j}\sqrt{\kappa_u\kappa_v}\in\bC^{MN}$ denotes the binary vector corresponding to the combined probing window with the angle support $\clU_{s,i} \times \clV_{s,j}$ from the beamforming codebook, which contains $1$ at the probed AoA-AoD components and $0$ elsewhere. An example of the probing geometry is illustrated in \figref{cluster} (b). Here we implicitly assume that the differences between the empirical and the statistical averages in \eqref{averageGamma}  \eqref{averageNoise} are also absorbed into the residual term $w_{s,i,j}$.

Following this procedure, over $T$ beacon slots, the UE obtains a total number of $M_{\text{RF}}N_{\text{RF}}T$ equations, which can be written in the form
\begin{align}\label{UE_equations}
\bfq=\bfB \cdot\vec{(\Gammam)} + \check{N}_cN_0R^x(0)\cdot \one + \bfw,
\end{align}
where the vector ${\bfq=[q_{1,1,1}, \dots q_{1,M_{\text{RF}},N_{\text{RF}}}, \dots, q_{T,M_{\text{RF}},N_{\text{RF}}}]^\transp}\in \bR^{M_{\text{RF}}N_{\text{RF}}T}$ consists of all $M_{\text{RF}}N_{\text{RF}}T$ measurements achieved as in (\ref{q_nocheck3}), ${\bfB=[\bfb_{1,1,1}, \dots, \bfb_{1,M_{\text{RF}},N_{\text{RF}}}, \dots, \bfb_{T,M_{\text{RF}},N_{\text{RF}}}]^\transp}\in \bR^{M_{\text{RF}}N_{\text{RF}}T\times MN}$ is uniquely defined by the pseudo-random beamforming codebook of the BS and the local beamforming codebook of the UE, and  $\bfw \in \bR^{M_{\text{RF}}N_{\text{RF}}T}$ denotes the residual fluctuations.

For later use, we define the SNR in each delay tap at the output of the matched filter \eqref{eq:j_out_beamspaceT1T2} as
	\begin{align}\label{snrPerchip}
	\snr^{y_{s,i,j}[k]}& \!=\! \frac{\pdim |R^{x^l}(0)|^2\sum_{l=1}^{L}\frac{\gamma_l+\eta_l}{1+\eta_l}\cdot \one_{\{kt_p=\tau_l\}}\!\cdot\! MN}{\bE[|z_{s,j}^c[k]|^2]\cdot \kappa_u\kappa_v}\nonumber\\
	&\overset{(a)}{\approx} \frac{\pdim|R^x(0)|^2\sum_{l=1}^{L}\frac{\gamma_l+\eta_l}{1+\eta_l}\cdot \one_{\{kt_p=\tau_l\}}\!\cdot\! MN}{N_0R^x(0)\cdot \kappa_u\kappa_v}\nonumber\\
	&=\frac{\ptot T_cN_c\sum_{l=1}^{L}\frac{\gamma_l+\eta_l}{1+\eta_l}\cdot \one_{\{kt_p=\tau_l\}}\cdot MN}{\kappa_u\kappa_vM_{\text{RF}}N_{\text{RF}}N_0}
	\end{align}
where $\one_{\{kt_p=\tau_l\}}$ is the indicator function, equaling to $1$ if $kt_p=\tau_l$ and $0$ otherwise, and where in $(a)$ we applied the approximation \eqref{R_app} by neglecting the matched-filtering loss due to the Doppler. It can be seen from \eqref{snrPerchip} that, large PN sequence duration (i.e., large $N_c$) implies an increase of the SNR in each sample \eqref{eq:j_out_beamspaceT1T2}, however, in order to avoid the effect of Doppler (e.g., in $(a)$), the PN sequence duration should not be longer than the channel coherence time $\Delta t_c$, i.e., $t_0=N_cT_c\leq \Delta t_c$, where $\Delta t_c$ is in general very small in mmWaves \cite{Rappaport2017lowrank}.

Note that, in our formulation, the effective observation over each beacon slot is actually the averaged energy $q_{s,i,j}$ defined by \eqref{q_nocheck}. For later analysis, we define the SNR in \eqref{q_nocheck} (or equivalently in \eqref{q_nocheck3}) as
\begin{align}\label{snrSum}
\snr^{q_{s,i,j}} &=\frac{\ptot T_cN_c\sum_{l=1}^{L}\frac{\gamma_l+\eta_l}{1+\eta_l}\cdot MN}{\check{N}_c\kappa_u\kappa_vM_{\text{RF}}N_{\text{RF}}N_0}\overset{(a)}{\approx} \frac{\ptot T_c\sum_{l=1}^{L}\frac{\gamma_l+\eta_l}{1+\eta_l}\cdot MN}{\kappa_u\kappa_vM_{\text{RF}}N_{\text{RF}}N_0}
\end{align}
where $(a)$ follows the fact that $\check{N}_c\approx N_c$, i.e., when $N_c$ is large enough (more precisely, when $t_0 \gg \Delta \tau_{\max}$), the relative difference between $\check{N}_c\geq N_c+\frac{\Delta \tau_{\max}}{T_c}$ and $N_c$ becomes negligible, since the delay spread in mmWave channels is  small \cite{Rappaport2014Capacity,Sayeed2014Sparse}.

To effectively capture the channel quality before beamforming (before BA),  we also define the SNR before beamforming (BBF) by
\begin{align}\label{snrBBF}
\snrbef =\frac{\ptot \sum_{l=1}^{L}\frac{\gamma_l+\eta_l}{1+\eta_l}}{N_0B}.
\end{align}
This is the SNR obtained when a single pilot stream ($M_{\text{RF}} = 1$)  is transmitted through a single BS antenna and is received in a single UE antenna  (isotropic transmission) via a single RF chain ($N_{\text{RF}} = 1$) over the whole bandwidth $B$. As mentioned before, one of the challenges of BA and in general communication at mmWaves is that the SNR before beamforming $\snrbef$ in \eqref{snrBBF} is typically very low.

\subsection{Path Strength Estimation via Non-Negative Least Squares}

We assume that the PSD $N_0$ of the AWGN channel is known for each UE \cite{SaeidBA2016}. In order to identify the AoA-AoD directions  of the strongest scatterers, the UE needs to estimate the $MN$-dim vector $\vec{(\Gammam)}$ from the  $M_{\text{RF}}N_{\text{RF}}T$-dim observation \eqref{UE_equations} in presence of the measurement noise $\bfw$, where in general, $MN$ is significantly larger than $M_{\text{RF}}N_{\text{RF}}T$. There are a great variety of algorithms to solve \eqref{UE_equations}. The key observation here is that $\Gammam$ is sparse (by sparse nature of mmWave channels) and non-negative (by second-order statistic construction of our scheme). As discussed in our previous work \cite{sxsBA2017}, recent results in CS show that when the underlying parameter $\Gammam$ is non-negative, the simple non-negative constrained {\em Least Squares} (LS) given by
\begin{align}\label{eq:NNLS}
\Gammam^\star=\argmin_{\Gammam \in \bR_+^{N\times M}} \|\bfB \cdot\vec{(\Gammam)} + \check{N}_cN_0R^x(0)\cdot \one - \bfq\|^2, 
\end{align}
is sufficient to impose the sparsity of the solution $\Gammam^\star$ \cite{slawski2013non,peter2018nnls}, without any need for an explicit
sparsity-promoting regularization term in the objective function as in the classical LASSO algorithm \cite{tibshirani1996regression}.
The (convex) optimization problem \eqref{eq:NNLS} is generally referred to as {\em Non-Negative Least Squares} (NNLS),
and has been well investigated in the literature. As discussed in \cite{slawski2013non}, NNLS implicitly performs $\ell_1$-regularization and promotes the sparsity of the resulting solution provided that the measurement matrix $\bfB$ satisfies the $\mathcal{M}^+$-criterion \cite{peter2018nnls}, i.e., there exits a vector $\bfd \in \bR_+^{M_{\text{RF}}N_{\text{RF}}T}$ such
that $\bfB^\transp \bfd >0$. In our case, this criterion can be simply interpreted as the fact that the set of $M_{\text{RF}}N_{\text{RF}}T$ measurement beam patterns should hit all the $MN$ AoA-AoD pairs at least once, which is almost fully satisfied in our scheme because of the finger-shaped beam patterns in each beacon slot, also because of the pseudo-random property of the designed beamforming codebook.

In terms of numerical implementations, the NNLS can be posed as an unconstrained  LS problem over the positive orthant
and can be solved  by several efficient techniques such as Gradient Projection, Primal-Dual techniques, etc., with an affordable computational complexity \cite{bertsekas2015convex}, generally significantly less than CS algorithms for problems of the same size and sparsity level. We refer to \cite{kim2010tackling, nguyen2015anti} for the recent progress on the numerical solution of
NNLS and  a discussion on other  related work in the literature.

\section{Performance Evaluation}\label{results}
We consider a system with $M=32$ antennas, $M_{\text{RF}}=3$ RF chains at the BS, and $N=32$ antennas, $N_{\text{RF}}=2$ RF chains at a generic UE. We assume a short preamble structure used in IEEE 802.11ad \cite{Time2017,ParameterPerahia2010}, where the beacon slot is of duration $t_0S=1.891\,\mu$s. The system is assumed to work at $f_0=70$ GHz, has a maximum available bandwidth of $B=1.76$ GHz, namely, each beacon slot amounts to more than $3200$ chips as in \cite{Time2017,AlkhateebTimeDomain2017}. We assume the channel contains $L=3$ links given by $(\gamma_l=1,\eta_1=100)$, $(\gamma_l=0.6,\eta_1=10)$ and $(\gamma_l=0.6,\eta_1=0)$,  where $\gamma_l$ denotes the scatterer strength, $\eta_l$ indicates the strength ratio between the LOS and the NLOS propagation as in \eqref{rice_fading}. Thus, the first scatterer can be roughly regarded as the LOS path, while the remaining scatterers represent the NLOS paths corresponding to two off-grid clusters. This is consistent with the practical mmWave MIMO channel measurements in \cite{Tim2018}, where the relative power level of the NLOS path is around $10$ dB lower than the desired LOS path. We assume that the relative speed $\Delta v_{l}$ for each path is around $0\sim8$ m$/$s. We announce a success if the location of the strongest component in $\Gammam^\star$ (see \eqref{eq:NNLS}) coincides with the LOS path \footnote{In the case that there is no LOS link, one can announce a success if the location of the strongest component in $\Gamma^\star$ coincides with the central AoA-AoD of the strongest scatterer cluster.}. 

\begin{figure}[t]
	\centerline{\includegraphics[width=8cm]{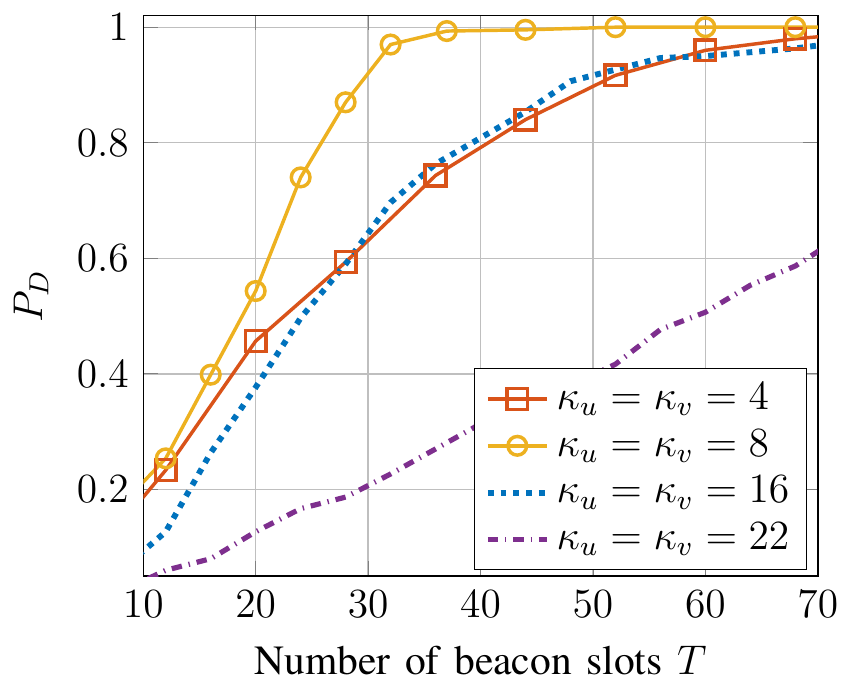}}
\caption{Detection probability $P_D$ of the proposed time-domain scheme with respect to different power spreading factors ($\kappa_u$, $\kappa_v$), where $M=N=32$, $M_{\text{RF}}=3$, $N_{\text{RF}}=2$, $B'=B$, $N_c=64$, $\snrbef=-14$ dB, the relative speed of the strongest path $\Delta v_{l^\star}=5$ m$/$s.}
\label{changeKvKu}
\end{figure}

In the following simulations \footnote{We will use \texttt{lsqnonneg.m} in\ \matlab  to solve the NNLS optimization problem in \eqref{eq:NNLS}.}, we evaluate the performance of our time-domain BA scheme according to two criteria: 
{i) We study the effect of various system parameters on the achieved BA probability. 
We also show the superiority of our proposed scheme in comparison with other recently proposed time-domain BA schemes \cite{AlkhateebTimeDomain2017,Time2017}; {ii) We consider the effectiveness of the BA scheme in the context of 
single-carrier modulation. This is obtained by computing upper and lower bounds on the ergodic achievable rate for the effective SISO 
channel after BA. These bounds show that BA yields essentially a flat channel even in the presence of multipath components. Therefore, 
single-carrier modulation without time-domain equalization  works very well.

\subsection{Success Probability of the Proposed BA Scheme}

{\bf Dependence on the beam spreading factors $(\kappa_u,\kappa_v)$}. As discussed in our previous work \cite{sxsBA2017,sxs2017Time}, the 
spatial spreading factors $(\kappa_u,\kappa_v)$ impose a trade-off between the angle coverage of the measuring matrix $\bfB$ and the SNR of received 
signal at the UE side. It can be seen from \figref{changeKvKu} that, increasing the spreading factor from $\kappa_u=\kappa_v=4$ to $\kappa_u=\kappa_v=8$ improves the performance. However, the performance keeps degrading when $(\kappa_u,\kappa_v)$ are increased to $\kappa_u=\kappa_v=16,22$.

%
%

\begin{figure}[t]
	\centerline{\includegraphics[width=15cm]{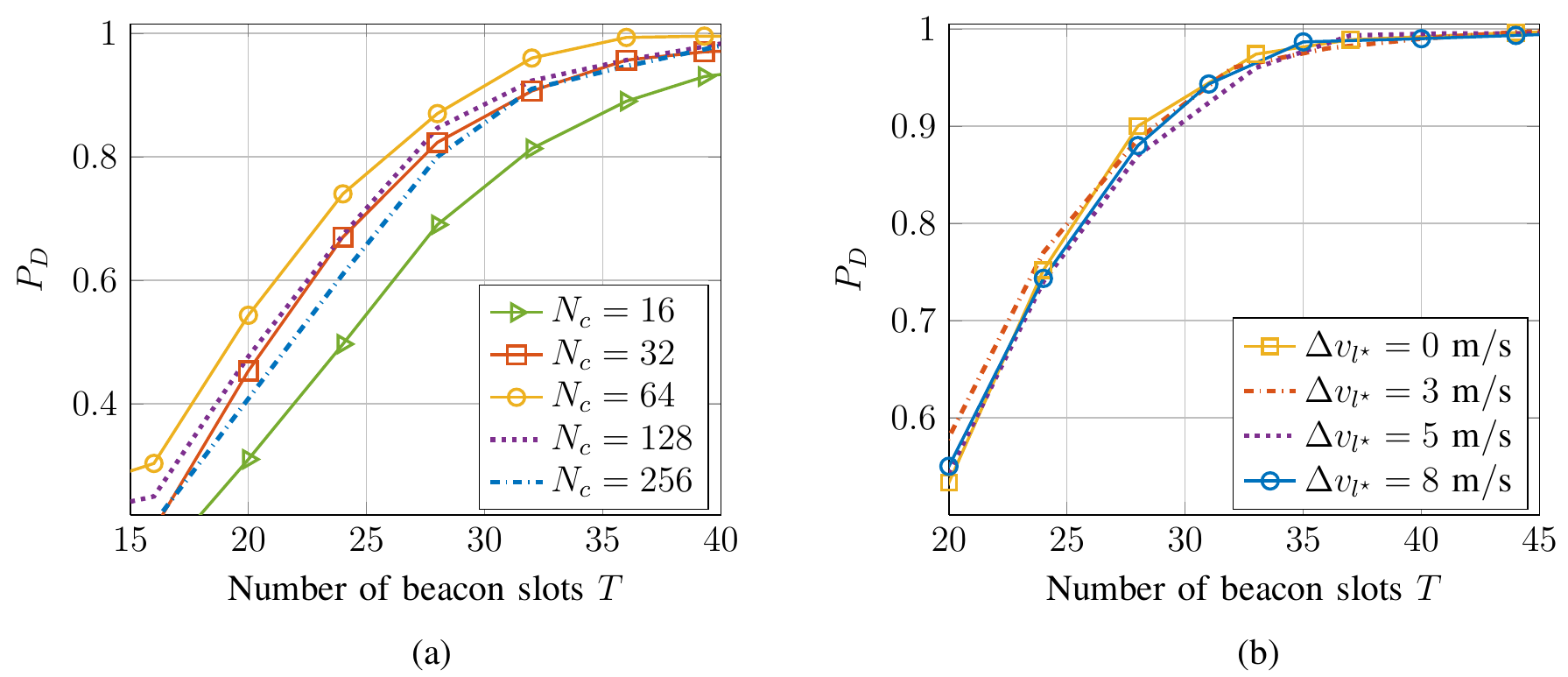}}
	\caption{ (a) Detection probability $P_D$ of the proposed time-domain scheme with respect to (a) different PN sequence lengths $N_c$, where the relative speed of the strongest path $\Delta v_{l^\star}=5$ m$/$s; (b) different relative speed values of the strongest path $\Delta v_{l^\star}$. In both cases, $M=N=32$, $M_{\text{RF}}=3$, $N_{\text{RF}}=2$, $\kappa_u=\kappa_v=8$, $B'=B$, $\snrbef=-14$ dB.}
	\label{changeNc-changeVV}
\end{figure}

{\bf Dependence on the PN sequence length $N_c$ and robustness to Doppler shifts}. In general, larger PN sequence length $N_c$ provides better correlation properties, such that different pilot streams can be well separated at the UE. However, increasing $N_c$ increases the whole duration $t_0=N_c T_c$ of the transmitted signal. Thus, because of the Doppler shift, the received PN sequence undergoes larger phase rotation of the chips. This rotation
degrades the PN sequence correlation property. This is illustrated in \figref{changeNc-changeVV} (a). As we can see, increasing the PN sequence length $N_c$ from $N_c=16$ to $N_c=32,64$ improves the performance of the proposed scheme. However, the performance degrades slightly when $N_c$ is increased to $N_c=128, 256$. Moreover, as shown in \figref{changeNc-changeVV} (b), the proposed scheme is highly insensitive to the Doppler spread between different multipath components. For example, varying the relative speed difference between the paths 
from $0$ to $8$ m$/$s,  the BA success probability remains virtually unchanged. 
This provides a significant advantage with respect to schemes based on OFDM signaling, 
which is known to be fragile to uncompensated Doppler shifts yielding inter-carrier interference.

{\bf Comparison with other time-domain methods.}
\figref{nnls_omp} compares the performance of our proposed scheme with a recently proposed time-domain 
approach \cite{Time2017,AlkhateebTimeDomain2017} based on the OMP CS technique. 
The approach in \cite{Time2017,AlkhateebTimeDomain2017} assumes that the channel vector coefficients 
remain constant over the whole training stage (in other words, it assumes a completely stationary situation with zero Doppler shifts).
It can be seen from \figref{nnls_omp} that the proposed scheme exhibits much more robust performance with respect to the channel  time-variations whereas the approach in \cite{Time2017,AlkhateebTimeDomain2017} fails when the channel is fast time-varying.

\begin{figure}[t]
	\centerline{\includegraphics[width=8cm]{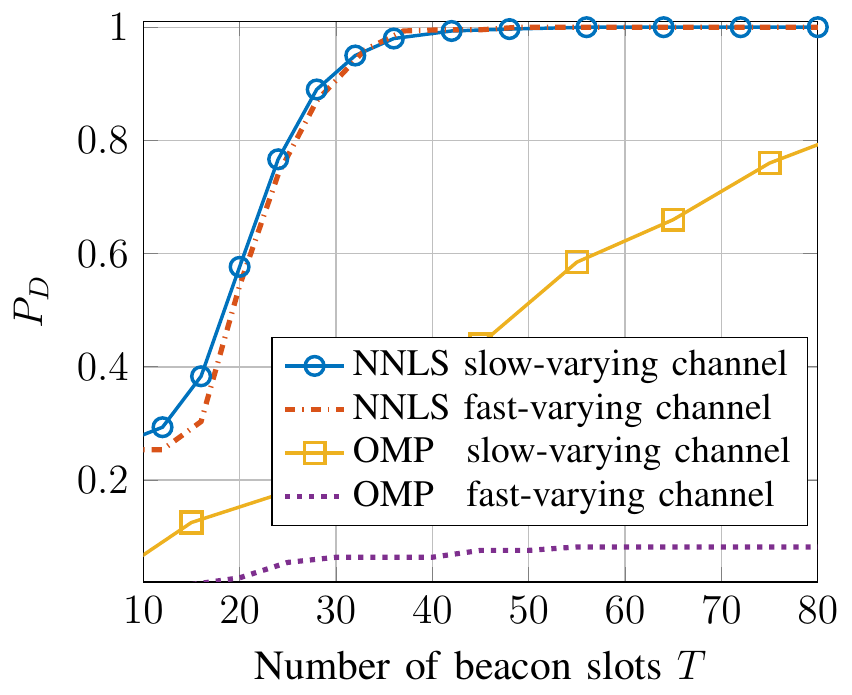}}
	\caption{{ Comparison of the proposed scheme based on NNLS with that in \cite{AlkhateebTimeDomain2017,Time2017} based on OMP  for both slow-varying and fast-varying channels, where $M=N=32$, $M_{\text{RF}}=3$, $N_{\text{RF}}=2$, $\kappa_u=\kappa_v=8$, $B'=B$, $N_c=64$, $\snrbef=-14$ dB}.}
	\label{nnls_omp}
\end{figure}

\subsection{Effectiveness of Single-Carrier Modulation}

 Assume that after a BA procedure as proposed in Section \ref{proposedscheme}, the strongest component in $\Gammam^\star$ corresponds to the $l^\star$-th scatterer between the BS and the UE. Hence, the estimated beamforming vectors for the data transmission are given by $\bfu_{l^\star} = \Fm_M \check{\bfu}_{l^\star}$ at the BS and $\bfv_{l^\star} = \Fm_N \check{\bfv}_{l^\star}$ at the UE respectively, where $ \check{\bfu}_{l^\star}\in\bC^M$ is an all-zero vector with a $1$ at the component corresponding to the AoD of the $l^\star$-th scatterer, and $ \check{\bfv}_{l^\star}\in\bC^N$ is an all-zero vector with a $1$ at the component corresponding to the AoA of the $l^\star$-th scatterer. We assume that in the downlink data transmission phase, the BS and the UE employ a single RF chain, therefore,  with a slight abuse of notation, we assume that transmitted waveform, consisting of $N_d$ information symbols, is given by $x(t)=\sum_{n=1}^{N_d}\sqrt{\ptot T_d}\cdot d_{n} p_{r}(t-nT_d)$, where $p_{r}(t)$ denotes the normalized band-limited pulse shaping filter (such as a raised cosine pulse), $T_d=1/B$ indicates the symbol transmission rate over the whole bandwidth $B$, and $\forall n\in[N_d]$, $d_{n}\in\{1,-1\}$ indicate the information symbols. From \eqref{receiveTT} and \eqref{eq:j_out}, the received signal after passing through the beamforming vectors $(\bfv_{l^\star},\bfu_{l^\star})$ is given by
\begin{align}\label{data_out}
\hat{y}(t)&=\bfv_{l^\star}^\herm\! \int\! \sfH(t,d\tau) x(t-\tau)\bfu_{l^\star}\! +\!z(t)\nonumber\\
&=\!\sum_{l=1}^L \sum_{n=1}^{N_d}C_{l}p_{r}(t-nT_d-\tau_l)e^{j2\pi (\check{\nu}_{l}+\nu_{l}nT_d)}\!+\!z(t),
\end{align}
where $C_{l} := \sqrt{\ptot T_d}\cdot\rho_{l}\bfv_{l^\star}^\herm\bfa_{\text{R}}(\phi_l) \bfa_{\text{T}}(\theta_l)^\herm\bfu_{l^\star}$. We assume that the UE performs a matched filter to $p_{r}(t)$, where the signal at the output of the matched filter can be written as
\begin{align}\label{data_filter}
y(t)|_{t=k T_d}&=\int \hat{y}(\tau)p_{r}^*(\tau-kT_d)d\tau\nonumber\\
&=\!\sum_{n=1}^{N_d}C_{l^\star}\varphi\left((k-n)T_d-\tau_{l^\star}\right)\!+\!\sum_{l\neq l^\star}\sum_{n=1}^{N_d}C_{l}\varphi\left((k-n)T_d-\tau_l\right)+z^c(t),
\end{align} 
where $z^c(t)$ denotes the noise at the output of the matched filter with a PSD $N_0$ (multiplied by $\int |p_{r}(t)|^2dt=1$), we define $\varphi(t-nT_d-\tau_l)|_{t=kT_d} = \int p_{r}((k-n)T_d-\tau_l)p_{r}^*(\tau-kT_d)d\tau\cdot e^{j2\pi (\check{\nu}_{l}+\nu_{l}nT_d)}$, $\forall l\in[L]$, where the amplitude of $\varphi\left((k-n)T_d-\tau_l\right)$ approximately equals to $1$ when $(k-n)T_d-\tau_l=0$. It can be seen from \eqref{data_filter} that, the first term denotes the desired signal, whereas the last two terms correspond the inter-symbol interference and noise.  we assume, for simplicity, that the delay tap of the strongest path $l^\star$ coincides with one of the sampling points at the UE side. Consequently, the ergodic achievable rate can be upper and lower bounded by \cite{caire2017Bound}
\begin{align}\label{ub_ergodic}
R^{ub^\star} =\bE\left[\log_2\left(1+\frac{\sum_{l=1}^{L}|C_{l}\varphi(\tau_{l^\star}-\tau_l)|^2}{N_0}\right)\right],
\end{align} 
\begin{align}\label{lb_ergodic}
R^{lb^\star}=\log_2\left(1+\frac{|\bE[C_{l^\star}\varphi(0)]|^2}{N_0+\var(C_{l^\star}\varphi(0))+\sum_{l\neq l^\star}\bE[|C_l\varphi(\tau_{l^\star}-\tau_l)|^2]}\right).
\end{align} 

The upper bound \eqref{ub_ergodic} is obtained via the \textit{Maximum Ratio Combining} for the case where all the delayed versions of the transmitted signal are separately observable (this is sometimes referred to as ``matched filter upper bound''). The lower bound is actually achieved
by a simple receiver that treats all the \textit{Inter-symbol Interference} (ISI) as a Gaussian noise.
As already noticed a few times in this paper, once BA is achieved, the effective channel angular spread is very small since essentially only 
the selected AoA/AoD path collects all the signal power. As a result, the channel consists of only a single dominant delay tap with a fixed Doppler shift, 
which can be well compensated by applying conventional timing and frequency synchronization
and channel estimation techniques.  Due to the Doppler compensation, the channel time-variations are significantly reduced after BA   \cite{HeathVariation2017},
such that $C_{l^\star}$ can be treated as an almost  deterministic channel gain with a very large amplitude $|\bE[C_{l^\star}\varphi(0)]|$ and a very  small variance $\var(C_{l^\star}\varphi(0))$.

\begin{figure}[t]
	\centerline{\includegraphics[width=8cm]{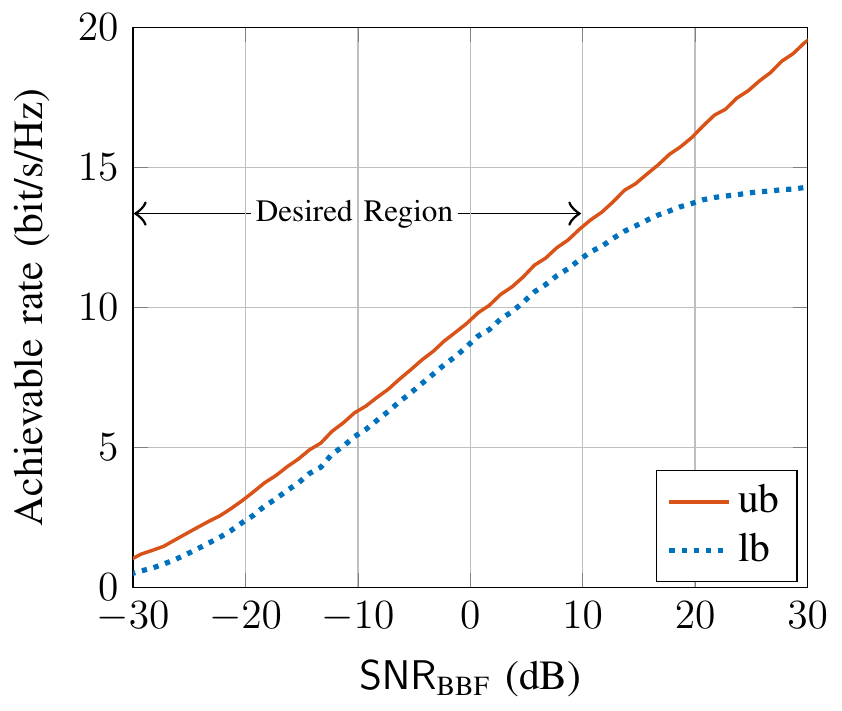}}
	\caption{The ergodic achievable rate after a successful {\em Beam Alignment} using the proposed time-domain scheme, where $M=N=32$, $B'=B$, $N_c=64$, the relative speed of the strongest path $\Delta v_{l^\star}=5$ m$/$s. }
	\label{rate}
\end{figure}

{\bf Ergodic achievable rate bounds}.
In \figref{rate}, we illustrate the lower and upper bounds on the  achievable ergodic rate  \eqref{ub_ergodic}  \eqref{lb_ergodic} as a function of $\snrbef$.
While it is clear that the lower bound is interference-limited while the upper bound is not, we notice that the gap between the bounds is quite small in the regime of low pre-beamforming SNR ($\snrbef < 10$\,dB), which is relevant in mmWave applications. At the same time, the achievable ergodic spectral efficiency	in this regime can be quite high. In particular, we remark here that the lower bound refers to the case of single-carrier transmission without any equalization. For example, focusing on a realistic spectral efficiency between 1 and 2 bit/s/Hz, we notice that single-carrier with the proposed BA scheme and no equalization (just standard	post-beamforming timing and frequency synchronization) achieves the relevant spectral efficiency in the range of $\snrbef$ between -30 and -20 dB, and suffers from a very small gap with respect to the best possible equalization (given by the upper bound).

\begin{figure}[t]
	\centerline{\includegraphics[width=8.5cm]{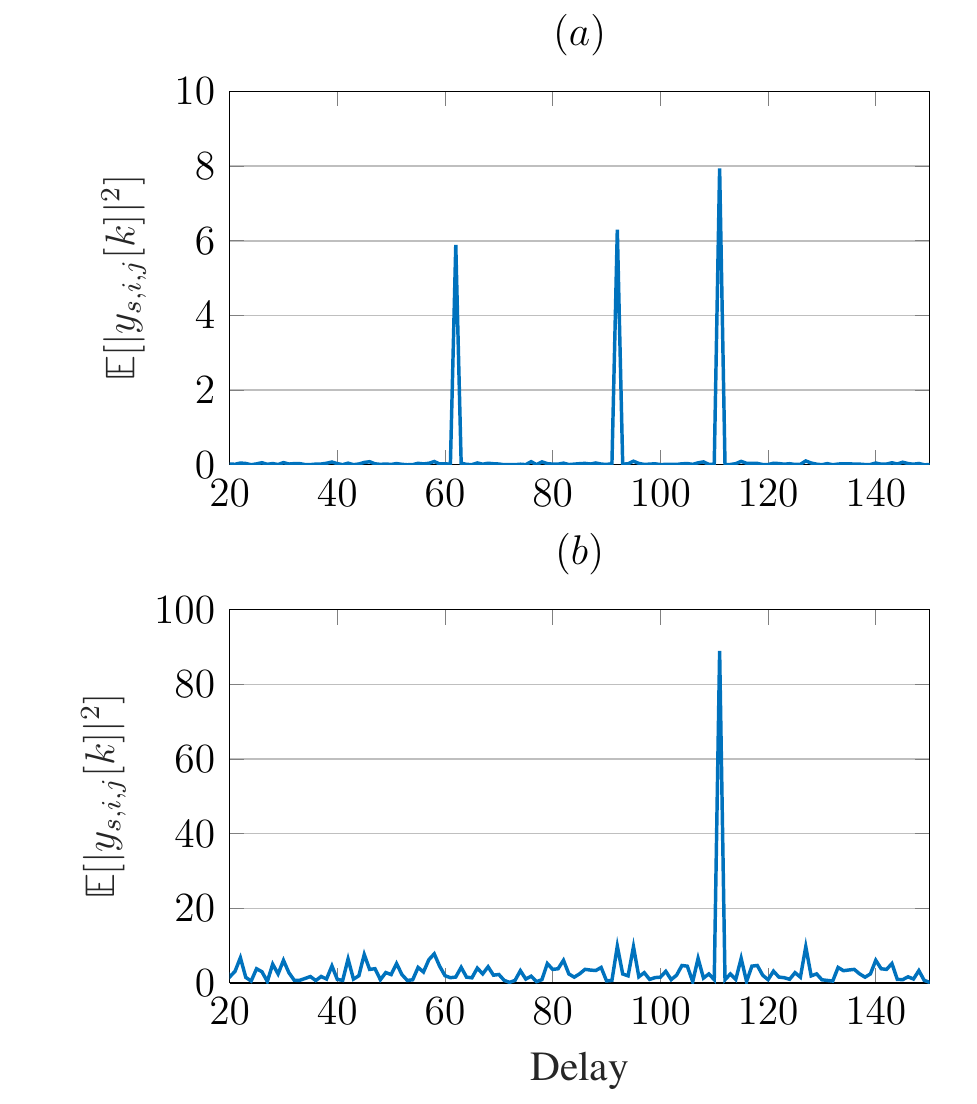}}
\caption{{Illustration of the {\em Power Delay Profile} (PDP) with multipath ($L=3$) channel in \eqref{eq:j_outFrom_i_sample}. $(a)$ Before {\em Beam Alignment}. $(b)$ After {\em Beam Alignment}}}
\label{DIP}
\end{figure}
{\bf Power Delay Profile (PDP) before and after Beam Alignment (BA)}. \figref{DIP} compares the average PDP of the mmWave channel with $L=3$ multipath components before and after BA. It can be seen from \figref{DIP} (a) that, before BA, the channel has a relatively large delay spread and is highly frequency selective. Moreover, since different multipath components are mixed with each other and since each one has its own delay and Doppler shift, the time-domain channel is highly time-varying. In contrast, as seen from \figref{DIP} (b), after BA, the channel effectively consists of a single multipath component, thus, it is almost flat in frequency. Also, note that in contrast with the former case where different multipath components were mixed with different Doppler frequencies, in the latter case the Doppler frequency of the single multipath component can be easily compensated by standard timing, frequency, and phase synchronization techniques at the receiver.

\section{Conclusion}

In this paper, we proposed a novel time-domain {\em Beam Alignment} (BA) scheme for 
mmWave MIMO systems with a HDA architecture. The proposed scheme is particularly suited for single-carrier multiuser mmWave communication, 
where each user has access to the whole bandwidth, and where all the users within the BS coverage can be trained simultaneously. 
We focused on the channel second-order statistics, incorporating both the random channel gains and Doppler shifts into the channel matrix to further capture the realistic features of mmWave channels. We applied the recently developed {\em Non-Negative Least Squares} (NNLS) technique to efficiently find the strongest path for each user. Simulation results showed that the proposed scheme incurs moderately low training overhead, achieves very good robustness to fast time-varying channels, and it is very robust to large Doppler shifts
among different multipath components. Furthermore, we have shown that the multipath channel after BA reduces essentially to a single giant tap
that collects almost all the signal energy. Hence, single-carrier signaling can perform very efficiently and requires just standard timing and frequency synchronization (that works well at high SNR after beamforming)
while it requires no time-domain equalization. This makes the proposed BA scheme together with single-carrier signaling a strong contender for future mmWave systems, especially in outdoor mobile scenarios.

\balance
{\footnotesize
\bibliographystyle{IEEEtran}
\bibliography{references}
}

\end{document}